\documentclass[aps,prc,twocolumn,groupedaddress]{revtex4}

\usepackage{epsfig}
\usepackage{graphicx}
\usepackage{fancyhdr}
\usepackage{pslatex}   
\usepackage{color}
\usepackage{amsmath, amsthm, amssymb} 

\newcommand{\Surd}{\mbox{$\surd$}}

\newcommand {\Alm} {{\mbox{$A_{l,m}$}}}
\newcommand {\AlmQ} {{\mbox{$A_{l,m}(Q)$}}}

\newcommand {\genbod} {{\sc genbod}~}
\newcommand {\pythia} {{\sc pythia}~}

\newcommand{\qo}{\mbox{$q_o$}}
\newcommand{\qs}{\mbox{$q_s$}}
\newcommand{\ql}{\mbox{$q_l$}}
\newcommand{\qoS}{\mbox{$q_o^2$}}
\newcommand{\qsS}{\mbox{$q_s^2$}}
\newcommand{\qlS}{\mbox{$q_l^2$}}

\newcommand{\RoS}{\mbox{$R_o^2$}}
\newcommand{\RsS}{\mbox{$R_s^2$}}
\newcommand{\RlS}{\mbox{$R_l^2$}}
\newcommand{\RolS}{\mbox{$R_{ol}^2$}}
\newcommand{\RslS}{\mbox{$R_{sl}^2$}}
\newcommand{\RosS}{\mbox{$R_{os}^2$}}

\newcommand{\YlmArg}[3]{\mbox{$Y_{l,m}\left({#1}\cos\theta,{#2}\phi{#3}\right)$}}

\newcommand{\Ylm}{\mbox{$Y_{l,m}$}}

\newcommand{\ImAlm}{\mbox{{\rm Im}[\Alm]}}
\newcommand{\ReAlm}{\mbox{{\rm Re}[\Alm]}}

\newcommand{\qcomp}[3]{\mbox{$\left({#1}\qo,{#2}\qo,{#3}\ql\right)$}}
\newcommand{\transform}[3]{\mbox{$\left(\qo,\qs,\ql\right)\rightarrow\left({#1}\qo,{#2}\qs,{#3}\ql\right)$}}

\newcommand{\prty}[1]{\mbox{$(-1)^{{#1}}$}}

\begin{document}


\title {Global Conservation Laws and Femtoscopy of Small Systems}


\author{Zbigniew Chaj\c{e}cki}
\email{chajecki@mps.ohio-state.edu}
\author{Mike Lisa}
\email{lisa@mps.ohio-state.edu}

\affiliation{ Department of Physics, Ohio State University,
1040 Physics Research Building, 191 West Woodruff Ave, Columbus, OH 43210, USA}



\begin{abstract}
It is increasingly important to understand, in detail, two-pion correlations
measured in $p+p$ and $d+A$ collisions.  In particular, one wishes to understand
the femtoscopic correlations, in order to compare to similar measurements in heavy
ion collisions.  However, in the low-multiplicity final states of these systems,
global conservation laws generate significant $N$-body correlations which project
onto the two-pion space in non-trivial ways and complicate the femtoscopic analysis.
We discuss a formalism to calculate and account for these correlations
in collisions dominated by a single particle species (e.g. pions).  We also
discuss effects on two-particle correlations between non-identical particles,
the understanding of which may be important in the study of femtoscopic space-time
asymmetries.
\keywords{proton collisions, femtoscopy, heavy ions, pion correlations, RHIC, LHC}
\end{abstract}

\pacs{25.75.-q, 25.75.Gz, 25.70.Pq}


\maketitle


\section{INTRODUCTION}

The unique and distinguishing feature of collisions between 
heavy ions is their large (relative to the confinement scale) size and the possibility to generate
{\it bulk} systems which may be described in thermodynamic terms, allowing to discuss the Equation
of State of strongly-interacting {\it matter}.  The primary evidence for the creation of bulk 
matter at the highest energies~\cite{Adams:2005dq,Adcox:2004mh,Back:2004je,Arsene:2004fa} is the existence of strong collective flow~\cite{Ollitrault:1992bk}.
The dominant feature of flow is the correlation between space and momentum which it generates; thus,
momentum-only observables such as $p_T$ spectra and azimuthal 
anisotropies~\cite{Adams:2005dq,Adcox:2004mh,Back:2004je,Arsene:2004fa} represent only an indirect
projection of the effect.  Femtoscopic measurements access space as a function of particle momentum,
thus providing the most direct probe of the most crucial feature of heavy ion collisions~\citep[c.f. e.g.][]{Lisa:2005dd}.
In particular, flow is manifest by a negative correlation between the ``HBT radius'' and the transverse
mass ($m_T$) of the particles~\cite{Pratt:1984su}.

Clearly, then, a detailed understanding of femtoscopic measurements in heavy ion collisions
is crucial to proving the existence of, or probing the nature of, the bulk system generated
in the collision.  It is in fact possible to quantitatively interpret both
the femtoscopic and momentum-only observations at RHIC-- in A+A collisions-- in consistent,
flow-dominated models of the system~\citep[e.g.][]{Retiere:2003kf}. 
All seems well.

However, it is important to understand the system size dependence of the apparent bulk behavior.
In this paper we discuss the possible complications in the comparison of  large and small systems.

\subsection{Hadron collisions as a reference to heavy ion collisions}
One of the most exciting signals at RHIC is the modification of the jet structure due to the 
bulk medium. In particular leading particle distributions~\cite{Adler:2003qi,Adams:2003kv}
and azimuthal correlations~\cite{Adler:2002tq} 
in A+A collisions are strongly suppressed relative to those from p+p collisions, at high $p_T$.

Especially since low $p_T$ observables directly reflect the bulk medium, 
it is reasonable to ask whether similar comparisons between Au+Au and p+p collisions 
reveal comparable  differences in the soft sector.

Common measurements of this type include total particle yields, $p_T$ spectra 
and azimuthal correlations ($v_2$). 
However, in each case, it is not clear if qualitative differences between small and large
systems are observed. 
Quantum number conservation laws in small systems can strongly affect particle yields~\citep[e.g.][]{Derreth:1985kk,Elze:1986db}.
However, modulo canonical suppression effects, equilibrium-based thermochemical fits to  yields 
from p+p collisions~\cite{Becattini:1997xx}
produce quite similar results to those from A+A collisions~\cite{Braun-Munzinger:2001ip}.
Blast-wave fits~\cite{Schnedermann:1993ws,Retiere:2003kf} to $p_T$ spectra from p+p collisions 
indicate sizable radial flow, though smaller than that seen in Au+Au collisions~\cite{Adams:2003xp}.
Intrinsic anisotropies from p+p collisions are considered as non-flow contribution to azimuthal 
correlations in A+A.
However, it is far from obvious that the finite values of $v_2$ from p+p collisions~\cite{Adams:2004bi}
do not arise from collective flow in the p+p collision {\it itself}.

A Quark-Gluon Plasma is usually considered a form of matter.  If a Quark-Gluon Plasma
were created in p+p collisions, as suggested by Bjorken~\cite{Bjorken:1982tu}, would it
display bulk properties?  A direct comparison of soft-sector observables in p+p and A+A
collisions is necessary, to address these issues~\cite{Weiner:2005gp,Troshin:2006hp}.
The imminent hadronic and heavy ion program at the LHC brings the relevance 
of such studies into strong relief. 

\subsection{Femtoscopy in p+p collisions}

More light might be brought to bear on this important question 
through femtoscopic measurements, which probe more directly the space-momentum correlations
generated in a collective system.

Though not as plentiful as in heavy ion collisions,
two-pion femtoscopic measurements are common in $e^++e^-$ or $p+p(\bar{p})$
collisions~\cite{Alexander:2003ug}.  In these collisions, too, ``HBT radii'' are observed
to fall with $m_T$.  Speculations of the physics behind this observation have included
Heisenberg uncertainty-based arguments, string-breaking phenomena, and temperature gradients;
an excellent overview may be found in~\cite{Kittel:2005fu}.
Preliminary measurements by the STAR Collaboration~\cite{Chajecki:2006sf} even suggest
that p+p collisions show similar collective behavior as A+A collisions. 
Distinguishing different physical mechanisms, however, requires a detailed understanding 
of the correlations themselves. 

Such an understanding is complicated by the clear observation of non-femtoscopic effects in 
two-pion correlations in small systems. 
For example, in A+A collisions the functional form (Gaussian or not) fitted to  
two-pion correlations\citep[e.g.][]{Lisa:2005dd} incorporates only femtoscopic effects.
Such fits for smaller 
systems~\citep[e.g.][]{Agababyan:1996rg} have required additional ad-hoc terms of non-femtoscopic nature.

\subsection{Non-femtoscopic correlations}

Femtoscopic correlations are those which depend directly on the two-particle coordinate-space separation
distribution~\citep[c.f.][]{Lisa:2005dd}. 
In general, such correlations are confined to low relative velocity. 
Non-femtoscopic correlations may arise from string fragmentation or global conservation laws, for example, and
there is no reason to expect that such correlations appear only in kinematic regimes
(e.g. ranges of relative momentum) different than
the femtoscopic ones.  Thus, separating those correlations may be a non-trivial exercise.

Non-femtoscopic correlations may arise from a variety of sources.
Jets will clearly induce momentum-space correlations between their fragmentation products.
While these effects may not be negligible, the low momentum of the pions under consideration ($p_T \sim 0.4$~GeV)
puts us squarely in the region in which factorization breaks down and the jet interpretation
becomes significantly murkier. 
In the kinematic region under consideration, string fragmentation may play a role; this is an
area for future study, though significant model-dependence will be present.
Collective bulk flow (e.g.
anisotropic elliptic flow) will generate $N$-body correlations which will project onto the two-body
space.  ``Clusters'' within events-- i.e. several independent particle-emitting sources-- may generate
additional structure~\cite{Gazdzicki:1992ri,Utyuzh:2007ct}; indeed, such clusters may be treated as ``large resonances''
with a many-body decay channel.
Each of these sources of non-femtoscopic correlations may play a greater or lesser role in a collision,
depending upon the physical scenario.
In this paper, we do not focus on these sources of correlation.

In this work, we focus on effects which must be at play in any physical system
-- Energy and Momentum Conservation-Induced Correlations
(EMCICs).
These global conservation laws provide an $N$-body constraint on the event, which projects down 
onto 2-body spaces and should become more pronounced at lower event multiplicity ($N$).


EMCIC effects on femtoscopic correlation functions
have been estimated~\citep[see Appendix C of ][]{Utyuzh:2007ct} recently in the context
of a numerical model of Bose-Einstein correlations from emitting cells, using a rough but fast numerical
algorithm to conserve energy-momentum.  An earlier study by Bertsch {\it et al.}~\cite{Bertsch:1993nx}
included energy-momentum conservation in an analytically-solvable model in the limit of one spatial dimension
and non-relativistic particles.  Both studies were confined to correlation functions in one dimension of
relative momentum.
In this paper, effects of EMCICs on three-dimensional correlation functions are studied in detail, using
numerical simulations and analytic expressions, both based on the restricted phase-space integral.

\subsection{Structure of this paper}

In Section~\ref{sec:genbod} we describe \genbod, an event generator which 
samples an inclusive momentum distribution 
subject only to constraints of energy and momentum conservation.
In Section~\ref{sec:SHD} we briefly discuss the harmonic representation
which provides a complete and natural characterization 
of the shape of the correlation function.
An extensive discussion of symmetry constraints on the harmonics is given in Appendix~\ref{app:symmetries}.

For the next three Sections, we focus on events in which only pions are emitted.
In Section~\ref{sec:GenBod} we use pion-only events from \genbod to illustrate the
effects of varying constraints, frames, and kinematic cuts, on EMCICs.
A method to calculate analytically (but using information from measured distributions) EMCICs is shown
in Section~\ref{sec:JYO}.  This leads to an ``experimentalist's formula,'' given in Section~\ref{sec:formula},
intended to disentangle
EMCICs from other (e.g. femtoscopic) correlations in the data.
The formula involves several approximations which may break down in reality; these
are discussed and effects are evaluated quantitatively.

In Section~\ref{sec:nonid} we discuss two effects that can complicate a direct
comparison of EMCICs for identical and non-identical particle correlations.  Several
interesting effects are observed, which might be important for the increasingly common studies of space-time
asymmetries.  We find that, for non-identical particles, the ``experimentalist's formula''
is only approximately applicable.
We summarize our discussion in Section~\ref{sec:summary}.

\section{Calculating events with energy and momentum conservation}
\label{sec:genbod}

To clearly understand the role of EMCICs, we would like to study events in which there is no other physics
involved besides the conservation laws.  Such a tool has been provided 40 years ago in the form of
the \genbod computer program~\cite[see ][ for an excellent write-up of the method and physics]{James:1968gu}
in the CERN library.  
Given a requested total multiplicity ($N$), a list of masses ($m_i$) of emitted particles, and a
total amount of energy ($E_{\rm tot}$) to distribute among them, \genbod returns an event of random
momenta (four-vectors $p_j$), subject only to the condition of energy and momentum conservation.  More importantly, it returns,
for each event, a weight proportional to the probability that the event will actually occur in nature.
Thus, it is a much different tool than, say, transport codes like RQMD~\cite{Sorge:1989vt},
in which each event returned may be treated as equally probable.

This weight is based on the phase-space integral $R_N$~\cite{Hagedorn:1963xx}
\begin{equation}
R_N = \int^{4N}\delta^4\left(P-\sum_{j=1}^N p_j \right)\prod_{i=1}^N\delta\left(p_i^2-m_i^2\right)d^4p_i ,
\end{equation}
where $P = \left(E_{\rm tot},\vec{0}\right)$ is the total momentum four-vector of the event.
$R_N$ figures dominantly in Fermi's statistical theory~\cite{Fermi:1950jd}, in which the probability of having
$N$ particles in the final state is proportional to $\bar{S}_N \cdot R_N$; here $\bar{S}_N$ is the phase-space-averaged $S$-function (or
matrix element) associated with the process generating the final state. 

In the limit for which momentum distribution is dominated by phase-space restrictions alone, $\bar{S}$ is a
constant, and the spectrum of a quantity $\alpha$ (say, an angle or transverse momentum) is
given by~\cite{Fermi:1950jd,Hagedorn:1963xx,James:1968gu}
\begin{equation}
\label{eq:alphaSpec}
f\left(\alpha\right) = \frac{d}{d\alpha}R_N .
\end{equation}
In the limit that $\alpha$ represents the ensemble of momenta constituting a given event, Equation~\ref{eq:alphaSpec}
returns the event weight.
See~\cite{James:1968gu} for a practical iterative prescription to calculate $R_N$ and the weights.

\begin{figure}[t]
{\centerline{\includegraphics[width=0.50\textwidth]{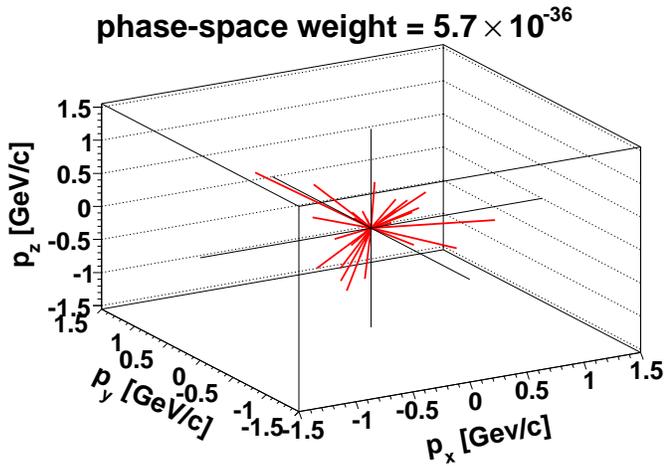}}}
\caption{(Color online) A high-probability multiplicity-30 event calculated by \genbod.
Lines correspond to particle momenta $p_x,p_y,p_z$.\label{fig:LikelyEvent}}
\end{figure}
\begin{figure}[t]
{\centerline{\includegraphics[width=0.50\textwidth]{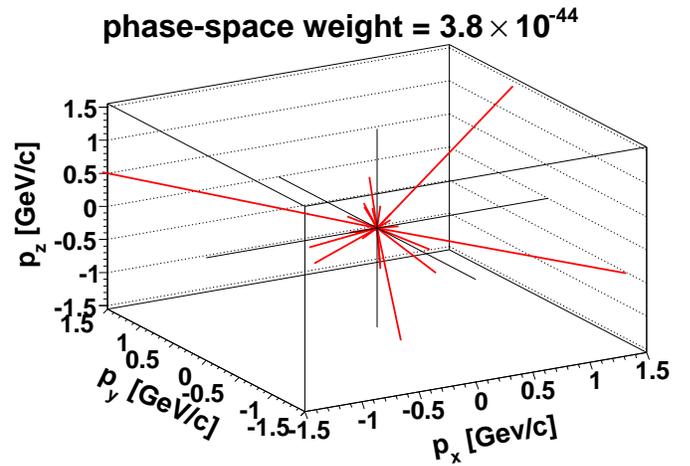}}}
\caption{(Color online) A low-probability multiplicity-30 event calculated by \genbod.
Lines correspond to particle momenta $p_x,p_y,p_z$.\label{fig:UnlikelyEvent}}
\end{figure}

We select (via Monte-Carlo) \genbod events according to their weight and run them through identical software
as used for experimental analysis.  
Fortunately, the code is fast, since one must calculate large statistics
from which to select.  
This is because the phase-space weights vary by large factors.  As a very extreme case,
Figures~\ref{fig:LikelyEvent} and~\ref{fig:UnlikelyEvent} show a likely and unlikely event, respectively,
for multiplicity $N=30$.  As one would expect, the ``rounder'' event is more likely, 
though one might be surprised
by the factor of a hundred million between the probabilities.

\section{SPHERICAL HARMONIC DECOMPOSITION OF CORRELATION FUNCTIONS}
\label{sec:SHD}

\begin{figure}[t]
{\centerline{\includegraphics[width=0.5\textwidth]{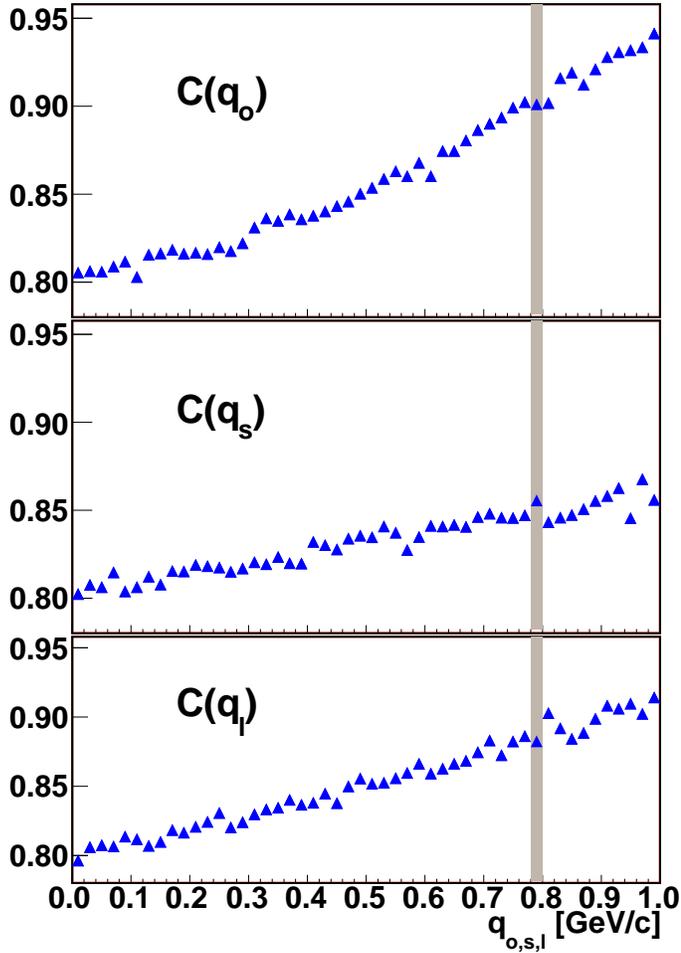}}}
\caption{(Color online) 1D projections of 3D correlation function calculated in LCMS frame for multiplicity-9 event calculated by \genbod.  \label{fig:GenBodN9prjs}}
\end{figure}

\begin{figure}[t]
{\centerline{\includegraphics[width=0.50\textwidth]{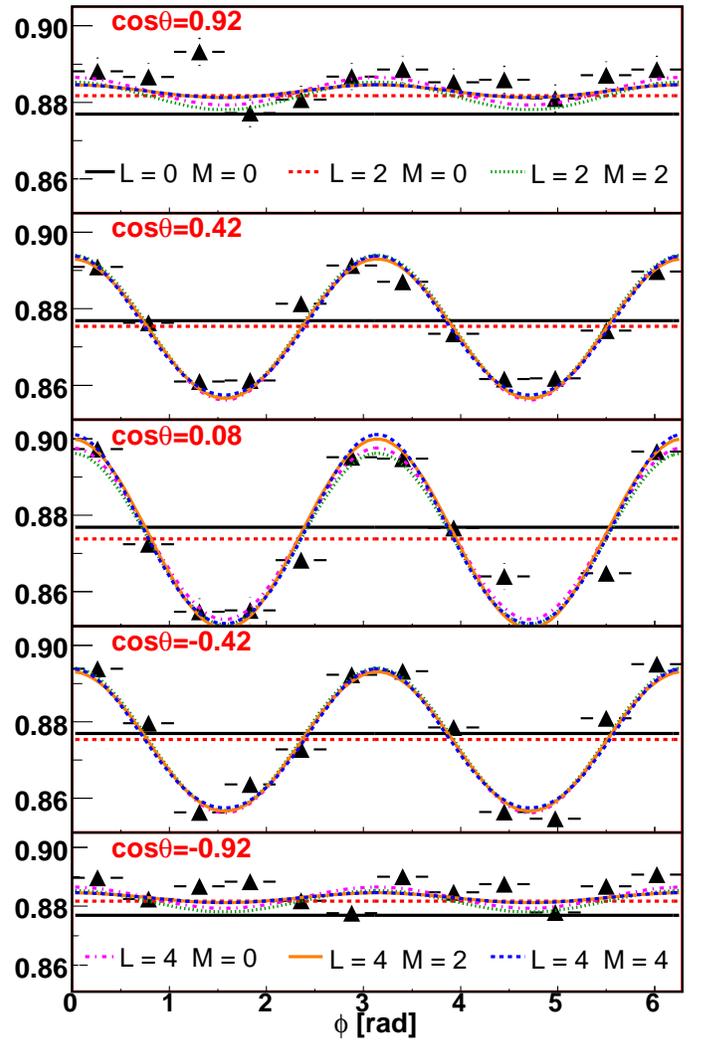}}}
\caption{(Color online) Correlation function  for the same data presented in Fig.~\ref{fig:GenBodN9prjs}
is shown at a fixed value of $Q=0.79~GeV/c$ (approximately indicated by the shaded region in
Fig.~\ref{fig:GenBodN9prjs}) as a function of $\phi$ for five bins in $\cos(\theta)$. Curves
represent the SHD components of various orders; see text for more details. \label{fig:qualityN9}}
\end{figure}

\begin{figure}[t]
{\centerline{\includegraphics[width=0.50\textwidth]{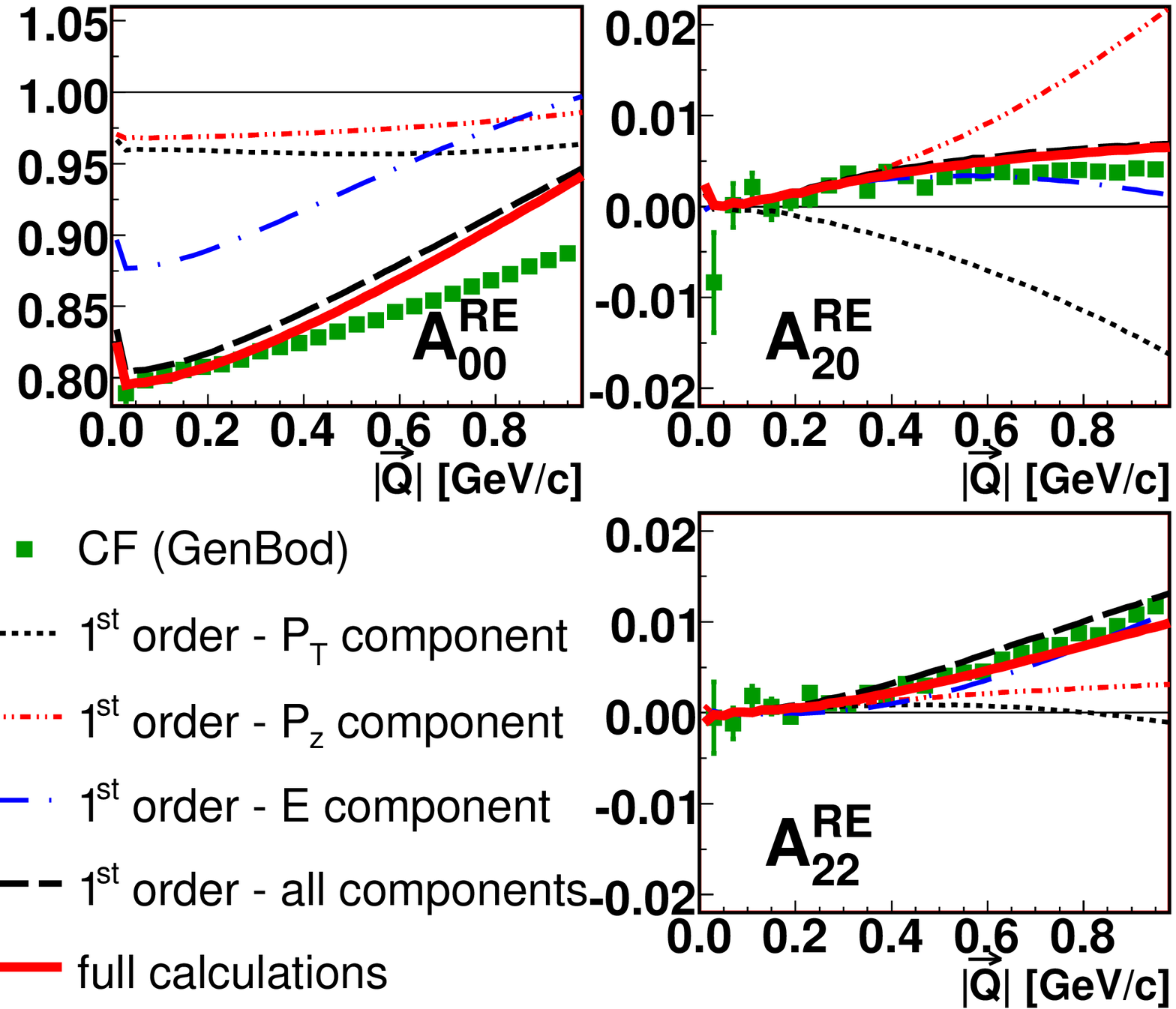}}}
\caption{(Color online) SHD coefficients for \genbod-generated events consisting of 9 pions having average kinetic energy per particle $\bar{K}=0.9$~GeV, as measured in the pair LCMS frame. 
No kinematic cuts were applied to data. Green squares are \Alm's from the \genbod events.  
Gold inverted triangles are the SHD coefficients of Equation~\ref{eq:JYOc2} for k=2.
Black circles, blue stars and red triangles are SHD coefficients of the first, second and third terms, respectively, of the
right side of Equation~\ref{eq:JYOc2firstorder}.  Open circles are SHD coefficients of the right side of Equation~\ref{eq:JYOc2firstorder}.
\label{fig:GenbodMult9LCMSK0p5}}
\end{figure}


Measurements of the space-time extent of a particle-emitting source at the femtometer scale is commonly done
by analyzing two-particle correlation functions $C\left(\vec{q}\right)$ as a function of
the relative momentum $\vec{q}$.  Experimentally, $C\left(\vec{q}\right)$ is the ratio of the $\vec{q}$ distribution
when both particles are measured in the same event, to the same distribution when the two particles come from
different events~\citep[c.f. e.g.][and references therein, for more details]{Lisa:2005dd}.
In this Section, we present correlation functions produced in exactly the manner in which experimental
ones are formed.  \genbod-generated events are selected according to the returned event weight.

Ideally, then, any structure remaining in this ratio reflects the correlation between particles
in the same event.  In the present study, these correlations come from energy and momentum conservation effects.

In this paper we will use the commonly used Bertsch-Pratt (``out-side-long'') 
decomposition of the relative momentum $\vec{q}$~\cite{Pratt:1986cc,Bertsch:1988db}, where $q_o$ is parallel to 
the transverse total momentum of the pair, $q_l$ is parallel to the beam direction 
and $q_s$ is perpendicular to those.


Usually the 3-D correlation functions are presented in 1-D Cartesian projections 
(or slices) along these axes (e.g. $q_o$) with the other q-components (e.g. $q_s$ and $q_l$) small.  
Such 1-D slices for \genbod calculations are presented on Figure \ref{fig:GenBodN9prjs}.
At asymptotically high relative momentum $|\vec{q}|$, femtoscopic contributions
to the the correlation function (those described by the Koonin-Pratt equation~\citep[discussed in][]{Lisa:2005dd})
must approach a constant value, usually normalized to unity, independent of the direction of $\vec{q}$.
Naturally there are no femtoscopic correlations in these events; correlations induced by the global
conservation laws are signaled by the non-unity value of the correlation function. 
The correlation function depends on the direction of $\vec{q}$ as well as $|\vec{q}|$.

However, one-dimensional projections represent a set of zero measure
of the 3-D correlation function and are thus 
a poor tool for exploring its detailed and potentially important structure.
In principle, one could visualize the full structure of the 3-D correlation function via a {\it series} of 
Cartesian projections in $q_i$ over different ranges in $q_{j,k}$, where $i \neq j \neq k$. 
This would, however, constitute a large number of figures, and relevant patterns which cut across
projections might not stand out. 

By exploiting symmetries in $\vec{q}$-space, 
the spherical harmonic decomposition (SHD)~\footnote{Danielewicz and Pratt have studied a similar decomposition
of the correlation function in terms of Cartesian Harmonics~\cite{Danielewicz:2005qh,Danielewicz:2006hi}.}
becomes a much more efficient representation 
which uses {\it all} of the data to show the shape of the correlation function.
Here, the spherical coordinates $\theta$, $\phi$, and $Q=|\vec{q}|$ relate to the Cartesian ones as
\begin{equation}
\label{eq:SHDcoords}
q_{o} = Q\sin{\theta}\cos{\phi} , \qquad 
q_{s} = Q\sin{\theta}\sin{\phi} , \qquad 
q_{l} = Q\cos{\theta} ,
\end{equation}
and we define harmonic moments \Alm's as
\begin{equation}
\label{eq:Alm}
\AlmQ \equiv \frac{1}{\sqrt{4\pi}} \int d\phi d(\cos\theta)  C\left(Q,\theta,\phi \right) Y_{l,m}\left(\theta,\phi\right).
\end{equation}

Usually, experimentally measured correlation functions are not continuous functions of $Q$, $\cos\theta$ and $\phi$, but
are constructed with bins of finite size.  In this case, Equation~\ref{eq:Alm} needs modification to account for finite-bin-size
effects.  For the experimental practitioner, we discuss one way to deal with this in Appendix~\ref{app:binning}.  For the remainder
of this manuscript,
we will assume these binning effects have been dealt with; i.e. we assume negligible bin size in $\cos\theta$ and $\phi$.

Symmetry constrains the number of relevant components.
For femtoscopic analyses of identical particles at midrapidity which integrate over reaction-plane orientation (i.e. almost all analyses to date), only 
real parts of \Alm's with even values of $l$ and $m$ do not vanish.
For the complete list of symmetries of \Alm's, see Appendix~\ref{app:symmetries}.
  Further, it is natural to expect that the statistical relevance of 
high-$l$ components is diminished.  

As an example, Fig.~\ref{fig:qualityN9} shows the calculated correlation function (the same as shown 
in Fig.~\ref{fig:GenBodN9prjs}) for one value of $Q$ as a function of $\cos\theta$ and $\phi$.
Also shown are curves representing SHD with increasingly higher order components.
In particular, the curves correspond to 
\begin{eqnarray}
\label{eq:recCF}
C_{L,M}(Q,\theta,\phi) &\equiv& \sqrt{4\pi} \left(\sum_{l=0}^{L-2} \sum_{m=-l}^{l} A_{l,m}(Q) Y^{*}_{l,m}(\theta,\phi) \right. \nonumber \\
&+&  \left. \sum_{m=-M}^{M} A_{L,m}(Q) Y^{*}_{L,m}(\theta,\phi)\right).
\end{eqnarray}
For example, the curve labeled as ``L=2 M=0'' contains $A_{0,0}$ (the constant term) 
and $A_{2,0}$ components.

Clearly, for this example, only the first few components are required to represent 
the structure of the correlation function. While a few higher-$l$ terms may be 
required in some cases, the number of relevant \Alm's is generically expected 
to be small. This is from general considerations of smoothness and, for experimental data, 
statistical issues.
Thus, by glancing at only a {\it few one-dimensional plots}, one views the {\it entire} correlation
structure in orthogonal components.   The number of plots is usually reduced further by symmetry constraints (c.f.~Appendix~\ref{app:symmetries}).

As an example, the first few \Alm's for the same \genbod calculations presented in 
this section, are plotted as a function of $Q$ in Fig.~\ref{fig:GenbodMult9LCMSK0p5}.
The odd-$l$ and -$m$ moments (not shown) vanish 
as required by symmetry (c.f. Appendix~\ref{app:symmetries}).


\section{EMCICS from \genbod}
\label{sec:GenBod}

\begin{figure}[t]
{\centerline{\includegraphics[width=0.50\textwidth]{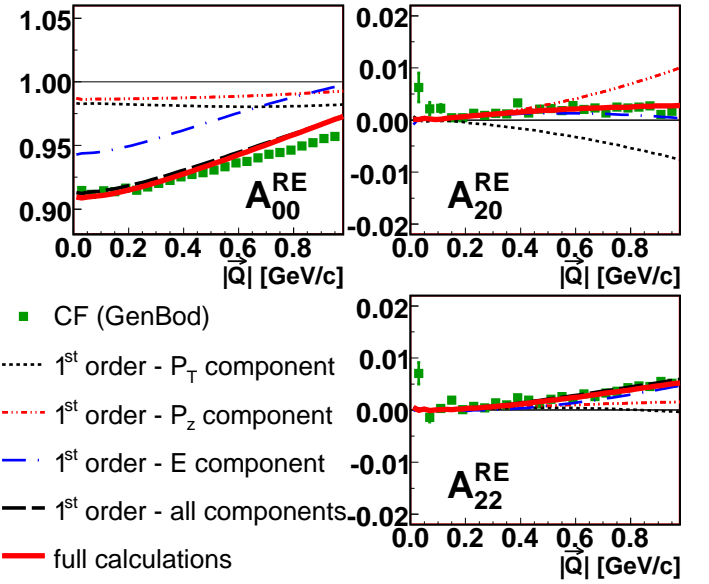}}}
\caption{(Color online) SHD coefficients for \genbod-generated events consisting of 18 pions having average kinetic energy per particle $\bar{K}=0.9$~GeV, as measured in the pair LCMS frame.
No kinematic cuts were applied to data. 
Green squares are \Alm's from the \genbod events.
Gold inverted triangles are the SHD coefficients of Equation~\ref{eq:JYOc2} for k=2.
Black circles, blue stars and red triangles are SHD coefficients of the first, second and third terms, respectively, of the
right side of Equation~\ref{eq:JYOc2firstorder}.  Open circles are SHD coefficients of the right side of Equation~\ref{eq:JYOc2firstorder}.
\label{fig:GenbodMult18LCMSK0p5}}
\end{figure}

\begin{figure}[t]
{\centerline{\includegraphics[width=0.50\textwidth]{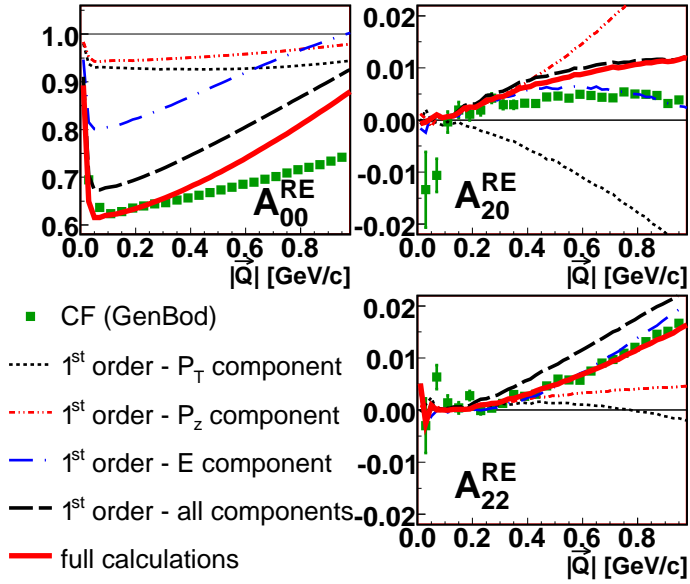}}}
\caption{(Color online) SHD coefficients for \genbod-generated events consisting of 6 pions having average kinetic energy per particle $\bar{K}=0.9$~GeV, as measured in the pair LCMS frame.
No kinematic cuts were applied to data.
Green squares are \Alm's from the \genbod events.
Gold inverted triangles are the SHD coefficients of Equation~\ref{eq:JYOc2} for k=2.
Black circles, blue stars and red triangles are SHD coefficients of the first, second and third terms, respectively, of the
right side of Equation~\ref{eq:JYOc2firstorder}.  Open circles are SHD coefficients of the right side of Equation~\ref{eq:JYOc2firstorder}.
\label{fig:GenbodMult6LCMSK0p5}}
\end{figure}

\begin{figure}[t]
{\centerline{\includegraphics[width=0.50\textwidth]{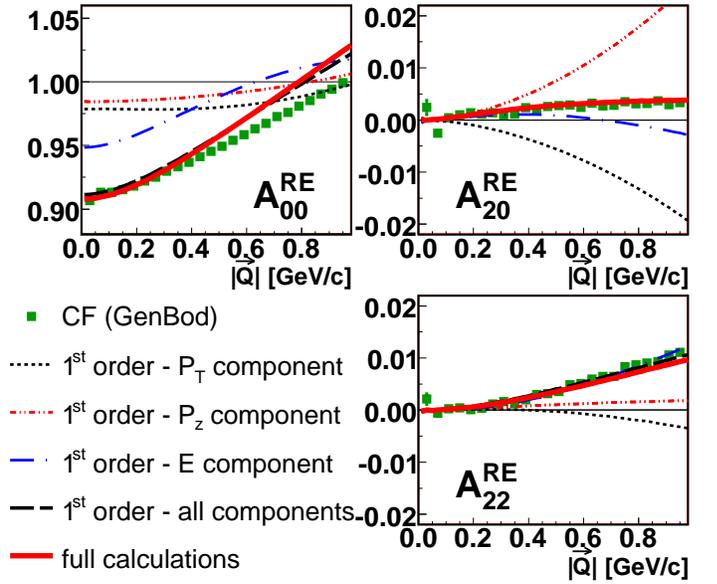}}}
\caption{(Color online) SHD coefficients for \genbod-generated events consisting of 18 pions having average kinetic energy per particle $\bar{K}=0.5$~GeV, as measured in the pair LCMS frame.
No kinematic cuts were applied to data.
Green squares are \Alm's from the \genbod events.
Gold inverted triangles are the SHD coefficients of Equation~\ref{eq:JYOc2} for k=2.
Black circles, blue stars and red triangles are SHD coefficients of the first, second and third terms, respectively, of the
right side of Equation~\ref{eq:JYOc2firstorder}.  Open circles are SHD coefficients of the right side of Equation~\ref{eq:JYOc2firstorder}.
\label{fig:GenbodMult18_en0.5_LCMSK0p5}}
\end{figure}

\begin{figure}[t]
{\centerline{\includegraphics[width=0.49\textwidth]{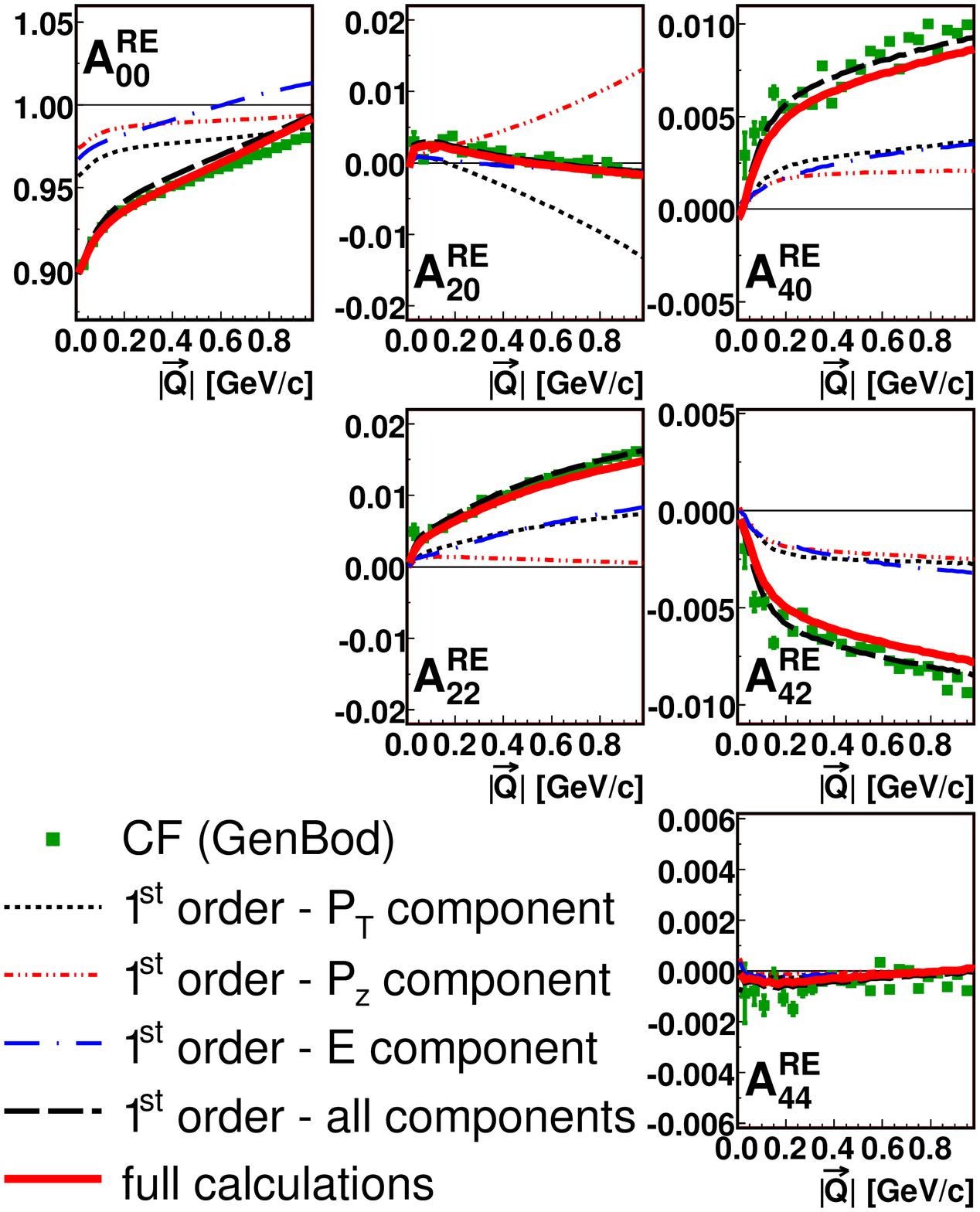}}}
\caption{(Color online) SHD coefficients for \genbod-generated events consisting of 18 pions having average kinetic energy per particle $\bar{K}=0.9$~GeV, as measured in the pair CMS frame. No kinematic cuts were applied to data.
Green squares are \Alm's from the \genbod events.
Gold inverted triangles are the SHD coefficients of Equation~\ref{eq:JYOc2} for k=2.
Black circles, blue stars and red triangles are SHD coefficients of the first, second and third terms, respectively, of the
right side of Equation~\ref{eq:JYOc2firstorder}.  Open circles are SHD coefficients of the right side of Equation~\ref{eq:JYOc2firstorder}.
\label{fig:GenbodMult18-noEtaCut}}
\end{figure}

\begin{figure}[t]
{\centerline{\includegraphics[width=0.49\textwidth]{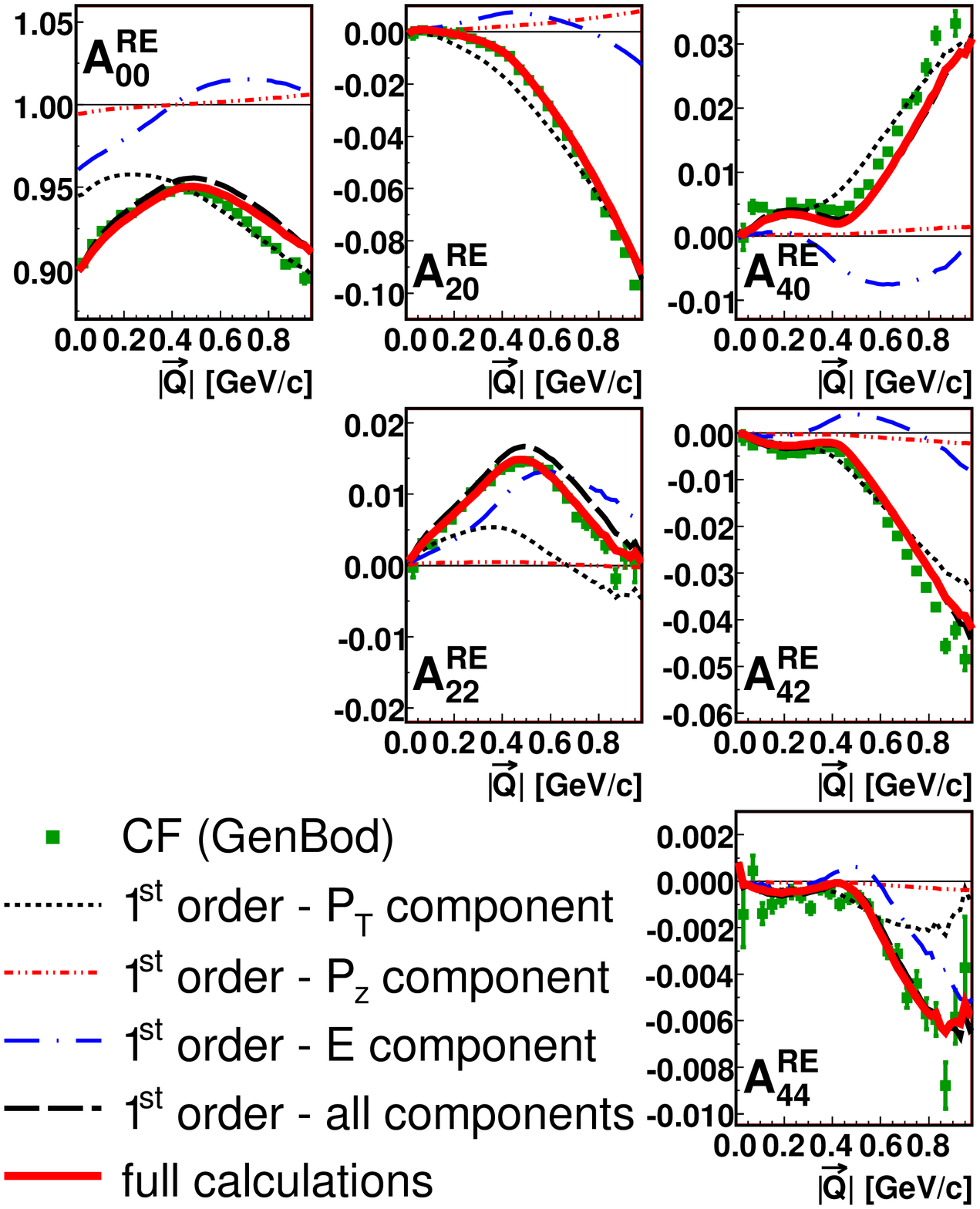}}}
\caption{(Color online) SHD coefficients for \genbod-generated events consisting of 18 pions having average kinetic energy per particle $\bar{K}=0.9$~GeV, as measured in the pair CMS frame. Only particles with $|\eta|<0.5$ used in the correlation function. Green squares are \Alm's from the \genbod events.  
Gold inverted triangles are the SHD coefficients of Equation~\ref{eq:JYOc2} for k=2.
Black circles, blue stars and red triangles are SHD coefficients of the first, second and third terms, respectively, of the
right side of Equation~\ref{eq:JYOc2firstorder}.  Open circles are SHD coefficients of the right side of Equation~\ref{eq:JYOc2firstorder}.
\label{fig:GenbodMult18-withEtaCut}}
\end{figure}

In this Section, we briefly discuss factors which affect the \Alm moments, using 
Figures~\ref{fig:GenbodMult9LCMSK0p5}-\ref{fig:GenbodMult18-withEtaCut}.  For the
present, we focus only on the green squares, labeled ``CF (GenBod),'' in those Figures.

Figures~\ref{fig:GenbodMult9LCMSK0p5},~\ref{fig:GenbodMult18LCMSK0p5}                      
and~\ref{fig:GenbodMult6LCMSK0p5} show the \Alm's calculated in LCMS frame \cite{Lisa:2005dd}
from \genbod events that have the same average kinetic energy per particle ($\bar{K}=0.9$~GeV)
but different multiplicity.
As expected, the strength of the EMCICs decreases with event multiplicity.
Similarly, for a given event multiplicity, one expects larger EMCICs when there is less available
energy.  As shown in Figures~\ref{fig:GenbodMult18LCMSK0p5} and~\ref{fig:GenbodMult18_en0.5_LCMSK0p5} 
for multiplicity-18 events, this is indeed the case.

Since the definition of the ``out,'' ``side'' and ``long'' directions-- and thus the angles $\theta$ and $\phi$--
depend on the frame of measurement, one expects the spherical harmonic coefficients \Alm to depend on reference
frame.  This is shown in Figures~\ref{fig:GenbodMult18LCMSK0p5} and~\ref{fig:GenbodMult18-noEtaCut}
for correlations measured in LCMS and pair CMS frames.

Less intuitive is the observation that the correlation strength depends also on kinematic cuts.
Figures~\ref{fig:GenbodMult18-noEtaCut}  and~\ref{fig:GenbodMult18-withEtaCut}
show the \Alm's calculated by \genbod
for 18-pion events without and with a selection of $|\eta|<0.5$, respectively.
(Note that this cut applies to the pions which are used in the analysis, {\it not}
to the set of particles for which energy and momentum is conserved; energy and momentum
is always conserved for the full event.)

Finally, we note two important and generic effects.  Firstly, EMCICs are present at all
values of $|\vec{Q}|$,
reminding us that we cannot (responsibly) ignore these effects in a femtoscopic analysis.
Secondly, in Figures~\ref{fig:GenbodMult18-noEtaCut}  and~\ref{fig:GenbodMult18-withEtaCut},
we have included \Alm components up to $l=4$.  Typically, $|A_{l+2,m}/\Alm|\sim 0.1$,
another reminder that characterization of the 3-dimensional correlation function requires
only a few harmonic components.

\section{ANALYTIC CALCULATION OF EMCICs}
\label{sec:JYO}

\vspace*{-3mm}

Even if EMCIC effects generated by \genbod ``resemble'' the experimental data, it is likely unwise
to use \genbod itself to correct the data for several reasons.  Firstly, there is strong sensitivity to
the (not completely measured) number and species-mix of {\it all} particles emitted in the event, including
neutrinos and possible magnetic monopoles (or, less exotically, particles escaping detector acceptance).
Secondly, there is strong sensitivity to the energy ``available'' in the event; it is not obvious that this
is $\sqrt{s_{NN}}$ of the collision.  Thirdly, EMCIC effects depend on the individual momenta $\vec{p}_1$ and
$\vec{p}_2$ of the particles entering the correlation function.  This will depend on acceptance, efficiency,
kinematic cuts
and, to a degree, the underlying single-particle phase-space.  (While correlation functions are insensitive to
the single-particle phase-space, the correlations which they measure may, in fact, depend on this phase-space,
due to physical effects.)

Thus, one would like to calculate EMCICs, based on the data itself.
In this Section, we begin by following arguments similar to those in
Refs~\cite{Danielewicz:1987in,Borghini:2000cm,Borghini:2003ur}
to obtain correction factors which implement EMCICs onto multi-particle distributions. 
 In the course
of the calculation, we make some simplifying approximations.  The derived expressions are then tested for accuracy
against the numerical \genbod simulations.  Finally, the expressions are
used to extract an ``experimentalist's formula'' discussed
in the next Section.

\subsection{Restricted phase-space corrections}
\label{sec:ItegralsGeneralization}

Danielewicz~\cite{Danielewicz:1987in}, and later Borghini, Dinh and Ollitrault~\cite{Borghini:2000cm},
considered  EMCIC-type effects on two-particle azimuthal correlations (quantified by $v_2$
and often used as a measure of elliptic flow~\cite{Ollitrault:1992bk}).
They focused mostly on transverse momentum ($\vec{P}_T$) conservation only, but Borghini later~\cite{Borghini:2003ur}
generalized to the case of an arbitrary number $D$ of independent (orthogonal) spatial dimensions and recently
considered momentum conservation effects on three-particle analyses of jet-like behavior~\cite{Borghini:2006yk}.

As we shall see below, for correlation functions used in femtoscopy,
the conservation of energy generates effects of similar magnitude as those due to conservation of
 \mbox{(three-)}momentum.  We deal only with on-shell particles, for which energy can{\it not} be treated as
independent of the momentum (as, say, $p_x$ would be largely independent of $p_y$).
Thus, unlike the above-mentioned works, we will explicitly begin with the 
more general multivariate central limit theorem.

We start with the case of interest-- $D=3$ spatial dimensions-- and conserve 3-momentum $\vec{p}$.
We implement energy conservation and on-shell constraints a bit
later.

We define~\footnote{Our use of symbols $f$ and $f_c$ follows the convention used in~\cite{Borghini:2000cm},
which is significantly different than-- if unfortunately similar-looking to-- that used in~\cite{Borghini:2003ur}
and~\cite{Borghini:2006yk}.}
\begin{equation}
\label{eq:DefineF}
f\left(\vec{p}_i\right) \equiv \frac{d^3N}{d\vec{p}^3_i}
\end{equation}
as the single-particle momentum distribution {\it un}affected by EMCICs.  This may be considered the unmeasured ``parent'' distribution.
Then, the $k-$particle distribution ($k$ less than the total multiplicity $N$) {\it including} EMCICs is
\begin{eqnarray}
\label{eq:BorghiniD3}
{f}_c\left(\vec{p}_1,...,\vec{p}_k\right) = \left(\prod_{i=1}^k {f}(\vec{p}_i)\right)\times  \qquad \qquad \qquad \qquad\nonumber \\
\frac{\int\left(\prod_{j=k+1}^N d^3\vec{p}_j {f}\left(\vec{p}_j\right)\right)\delta^3\left(\sum_{i=1}^N \vec{p}_i\right)}
{\int\left(\prod_{j=1}^N d^3\vec{p}_j {f}\left(\vec{p}_j\right)\right)\delta^3\left(\sum_{i=1}^N \vec{p}_i\right)} .
\end{eqnarray}
Note the difference between numerator and denominator in the starting value of the index $j$ on the product.

We implement total energy conservation
$\sum E_i = \sqrt{s}$, by
replacing $\delta^3\left(\sum_{i=1}^N \vec{p}_{i}\right) \rightarrow \delta^4\left(\sum_{i=1}^N p_i-P\right)$
in Equation~\ref{eq:BorghiniD3}.  Here, $P = \left(\sqrt{s},\vec{0}\right)$ is the total energy-momentum of the event, and 
$p_{0,i} = E_i = \sqrt{\vec{p}_i^2+m_i^2}$ 
is the energy of the on-shell particle.

We denote Lorentz-invariant distributions as
\begin{equation}
\label{eq:DefineTildeF}
\tilde{f}(p_i) \equiv 2E_i\frac{d^3N}{d\vec{p}_i^3} = 2E_i f(p_i) 
\end{equation}
and rewrite Equation~\ref{eq:BorghiniD3} as
\begin{eqnarray}
\label{eq:JYO1}
&&\tilde{f}_c\left(p_1,...,p_k\right) 
 = 
\left(\prod_{i=1}^k \tilde{f}(p_i)\right)\times  \qquad \qquad \qquad \qquad\nonumber \\
& & \frac{\int\left(\prod_{j=k+1}^N \frac{d^3\vec{p}_j}{E_j} \tilde{f}\left(\vec{p}_j\right)\right)\delta^4\left(\sum_{i=1}^N p_i - P\right)}
{\int\left(\prod_{j=1}^N \frac{d^3\vec{p}_j}{E_j} \tilde{f}\left(\vec{p}_j\right)\right)\delta^4\left(\sum_{i=1}^N p_i - P\right)} \nonumber \\
& = &
\left(\prod_{i=1}^k \tilde{f}(p_i)\right)\times  \qquad \qquad \qquad \qquad\nonumber \\
& & \frac{\int\left(\prod_{j=k+1}^N d^4p_j \delta\left(p^2_j-m^2_j\right)\tilde{f}\left(p_j\right)\right)\delta^4\left(\sum_{i=1}^N p_i-P\right)}
{\int\left(\prod_{j=1}^N d^4p_j \delta\left(p^2_j-m^2_j\right)\tilde{f}\left(p_j\right)\right)\delta^4\left(\sum_{i=1}^N p_i-P\right)}  \nonumber \\
& = &
\left(\prod_{i=1}^k \tilde{f}(p_i)\right)\times  \qquad \qquad \qquad \qquad\nonumber \\
& & \frac{\int\left(\prod_{j=k+1}^N d^4p_j g\left(p_j\right)\right)\delta^4\left(\sum_{i=1}^N p_i-P\right)}
{\int\left(\prod_{j=1}^N d^4p_j g\left(p_j\right)\right)\delta^4\left(\sum_{i=1}^N p_i-P\right)} .
\end{eqnarray}
Thus, we arrive at an integral over four independent variables, in which the integrand 
function $g(p)$ is ``highly peaked'' and with strong correlations in the 4-d $p-$space.

According to Equation~\ref{eq:JYO1}, the $k$-body momentum distribution, including EMCICs, is the
$k$-body distribution {\it not} affected by EMCICs-- i.e. just an uncorrelated product of single-particle distributions--
multiplied by a ``correction factor'' which enforces the EMCIC.  The numerator of this factor counts the number of configurations
in which the remaining $N-k$ on-shell particles conspire to conserve total energy and momentum,
and the denominator normalizes the distribution.

\subsection{Application of the Central Limit Theorem}
\label{sec:CLT}

To arrive at a useful result, we argue along lines similar to those of~\cite{Danielewicz:1987in,Borghini:2000cm,Borghini:2003ur}.
The distribution of 
a large number $M$ of uncorrelated momenta $W = \sum_{i=1}^M  p_i$ is, by the Central Limit Theorem, a multivariate normal distribution
\begin{eqnarray}
\label{eq:CLT_FmGeneral}
F_M\left(W\right) & \equiv &  \int \left(\prod_{i=1}^M  d^4 p_i g(p_i)\right) \delta^4\left(\sum_{i=1}^M p_i-W\right)  \\
& = & \sqrt{\frac{|B|}{\left(2\pi \right)^4}} \times \nonumber \\
&& \exp\left(-\frac{1}{2} \left(W^{\mu} - \langle P^{\mu} \rangle \right) B_{\mu \nu} \left(W^{\nu} - \langle P^{\nu} \rangle \right)  \right). \nonumber
\end{eqnarray}
Here, the average of the sum of 4-momenta is simply related to the single-particle average of the 4-momenta as
\begin{eqnarray}
\label{eq:aveP}
\langle P^{\mu} \rangle &=& \sum_{i=1}^{M} \langle p_i^{\mu} \rangle = M  \langle p_i^{\mu} \rangle,
\end{eqnarray}
where
\begin{eqnarray}
\label{eq:defineRMS}
\langle p^n_\mu \rangle \equiv \frac{\int d^4p g(p) \cdot p_\mu^n}{\int d^4p g(p)}, \nonumber \\
\langle p_\mu p_\nu \rangle \equiv \frac{\int d^4p g(p)  p_\mu p_\nu}{\int d^4p g(p)} .
\end{eqnarray}
Finally, in Equation~\ref{eq:CLT_FmGeneral},
 $|B|$ denotes the determinant of the matrix $B$.  Up to a factor of $M$, $B$ is the inverse of the covariance matrix of the distribution $g(p)$:
\begin{eqnarray}
B_{\mu \nu} &=& \frac{1}{M} b_{\mu \nu},
\end{eqnarray}
\begin{eqnarray}
\left(b^{-1} \right)_{\mu \nu} &=&  \langle p_{\mu} p_{\nu} \rangle - \langle p_{\mu} \rangle \langle p_{\nu} \rangle .
\label{eq:bmatrix}
\end{eqnarray}

We can now apply the CLT by recognizing the integral in the numerator in Equation~\ref{eq:JYO1} as the distribution of 
 $N-k$ momenta $\sum_{j=k+1}^N p_j = P - \sum_{j=1}^k p_j$
so that for ``large enough'' $N-k$, we find
\begin{eqnarray}
\label{eq:JYO2}
&&\tilde{f}_c\left(p_1,...,p_k\right) =   \left(\prod_{i=1}^k \tilde{f}(p_i)\right) \frac{F_{N-k} \Big(P - \sum_{i=1}^k p_i \Big)}{F_N(P)} \nonumber \\    
&&= \left(\prod_{i=1}^k \tilde{f}(p_i)\right)\cdot 
\left(\frac{N}{N-k}\right)^2\times \\
&&\exp
\left[- 
\left( \sum_{i=1}^k \left(p_i^{\mu} - \langle p^{\mu} \rangle\right)  \right) 
\frac{b_{\mu \nu}}{2\left(N-k\right)}
\left( \sum_{i=1}^k \left(p_i^{\nu} - \langle p^{\nu} \rangle\right)  \right) 
\right]. \nonumber
\end{eqnarray}

It is appropriate at this point to repeat the two approximations we have employed up to now.
The first assumption, always important in using the
CLT, is that $N-k$ is sufficiently large; recall that $N$ is the total multiplicity and $k$ is the order of
the correlation being calculated ($k=2$ for two-particle correlations).
Secondly, we have implicitly assumed that all particles in the system are governed by the {\it same} single-particle
distribution $g(p)$.  Strictly speaking, then, the system must consist of particles all of the same mass, and if there
are several species with the same mass (say, $\pi^-$ and $\pi^+$), they must furthermore have the same momentum
distribution.  This is at best an approximation for hadron or ion collisions, in which other particles contribute
to the pion-dominated final state.

\subsection{Observable EMCIC effects}

Even the single-particle momentum distribution is affected by EMCICs:
\begin{eqnarray}
\label{eq:JY0singleGeneral}
\tilde{f}_c\left(p_i\right) &=& \tilde{f}\left(p_i\right) \cdot \left(\frac{N}{N-1}\right)^2 \times \nonumber \\
&& \exp\left[-
 \left(p^{\mu} - \langle p^{\mu} \rangle\right)  
\frac{b_{\mu \nu}}{2\left(N-k\right)}
  \left(p^{\nu} - \langle p^{\nu} \rangle\right) 
\right]
\end{eqnarray}
The product of such a single particle distribution forms the denominator of the 
$k$-particle correlation function
\begin{eqnarray}
\label{eq:JYOc2General}
C\left(p_1,...,p_k\right)  \equiv 
  \frac{\tilde{f}_c\left(p_1,...,p_k\right)}{\tilde{f}_c\left(p_1\right)\cdots\tilde{f}_c\left(p_k\right)} 
   = 
  \frac{\left(\frac{N}{N-k}\right)^2}{\left(\frac{N}{N-1}\right)^{2k}} \times \qquad \\
  \frac{\exp\left[
\frac{-1}{2(N-k)}
 \sum_{i,j=1}^k \left(p_i^{\mu} - \langle p^{\mu} \rangle\right) 
b_{\mu \nu}
 \left(p_j^{\nu} - \langle p^{\nu} \rangle\right)  
   \right]
}
{
\exp
\left[\frac{-1}{2(N-1)}
 \sum_{i=1}^k \left(p_i^{\mu} - \langle p^{\mu} \rangle\right) 
b_{\mu \nu}
 \left(p_i^{\nu} - \langle p^{\nu} \rangle\right)  
\right]}  \nonumber
\end{eqnarray}

In this paper we concentrate on correlation functions in $q_{out},q_{side}$ and $q_{long}$, as is done in femtoscopic studies. However, the two-particle correlation function in relative azimuthal angle, which probes elliptic flow,
may also contain EMCIC contributions through Equation~\ref{eq:JYOc2General}.
These effects turn out to be small and are discussed in Appendix~\ref{app:V2}.
 
To first order in $1/N$, the two-particle correlation function becomes
\begin{eqnarray}
\label{eq:JYOc2firstorderGeneral}
C(p_1,p_2)=1- 
     \frac{1}{N}
\left( p_1^{\mu} - \langle p^{\mu} \rangle \right)
b_{\mu \nu}
\left( p_2^{\nu} - \langle p^{\nu} \rangle \right) .
\end{eqnarray}

The multivariate CLT used in Section~\ref{sec:CLT} accounts for correlations between vector components via
the covariance matrix $b^{-1}$ (Eq.~\ref{eq:bmatrix}) which has, in general, 10 nonvanishing elements.  The
average vector $P$ (Eq.~\ref{eq:aveP}) has in general 4 nonvanishing elements.  We
now reduce these numbers significantly by considering the specific case of our interest.

Firstly, we choose to work in the global center-of-momentum frame, so that
\begin{equation}
\label{eq:comFrame}
\langle p^\mu\rangle = \delta_{\mu,0}\langle E \rangle
\end{equation}

As for the correlations, we are interested in signals generated by EMCICs alone, not, for example, dynamical correlations due
to flow.  Neglecting elliptic flow (azimuthal anisotropies in the parent distribution~\cite{Ollitrault:1992bk,Poskanzer:1998yz})  implies
\begin{equation}
\label{eq:NoEllipticFlow}
\left(b^{-1}\right)_{1,2} = \langle p_x p_y\rangle = 0 .
\end{equation}
The same approach was adopted in earlier work~\cite{Danielewicz:1987in,Borghini:2000cm,Borghini:2006yk}.
Similarly, we assume no dynamical correlations due to directed flow~\cite{Poskanzer:1998yz}, implying
\begin{equation}
\label{eq:NoDirectedFlow}
\left(b^{-1}\right)_{1,3} = \left(b^{-1}\right)_{2,3} = 0 .
\end{equation}

The on-shell constraint generates an unavoidable dependence between energy and 3-momentum components.  However, in the
CLT limit, only the second moment (covariance) comes into play, and this vanishes.  For $i\neq 0$,
\begin{align}
\left(b^{-1}\right)_{0,i} = \langle E p_i \rangle - \langle E\rangle\langle p_i\rangle = \langle E p_i\rangle \\
= \frac{\int dE \int d^3\vec{p} \cdot Eg\left(p\right) \cdot p_i}{\int dE \int d^3\vec{p}\cdot g\left(p\right)} = 0 . \notag
\end{align}
In the last step, we recognize that $p_i$ is an odd function of momentum, whereas $E$ and $g$ are even.

In this scenario of interest, then, $b$ is diagonal, and Equations~\ref{eq:JY0singleGeneral} becomes
\begin{align}
\label{eq:JY0single}
& \tilde{f}_c\left(p_i\right) = \tilde{f}\left(p_i\right) \cdot \left(\frac{N}{N-1}\right)^2 \times \\
& \exp\left[-\frac{1}{2(N-1)}\left(
\frac{p^2_{i,x}}{\langle p_x^2 \rangle}+\frac{p^2_{i,y}}{\langle p_y^2 \rangle}+\frac{p^2_{i,z}}{\langle p_z^2 \rangle}
+\frac{\left(E_i-\langle E \rangle\right)^2}{\langle E^2 \rangle -\langle E \rangle^2}\right)\right] . \notag
\end{align}

Similarly, Equation~\ref{eq:JYOc2General} becomes
\begin{eqnarray}
\label{eq:JYOc2}
C\left(p_1,...,p_k\right)  \equiv 
  \frac{\tilde{f}_c\left(p_1,...,p_k\right)}{\tilde{f}_c\left(p_1\right)\cdots\tilde{f}_c\left(p_k\right)} 
   = 
  \frac{\left(\frac{N}{N-k}\right)^2}{\left(\frac{N}{N-1}\right)^{2k}} \times \qquad \\
  \frac{\exp\left[\frac{-1}{2(N-k)} \left\{
     \sum_{\mu=1}^{3}\left(\frac{\left(\sum_{i=1}^k p_{i,\mu}^2\right)^2}{\langle p_\mu^2\rangle}\right)
     +
     \frac{\left(\sum_{1}^k \left( E_i - \langle E\rangle\right)\right)^2}{\langle E^2 \rangle - \langle E \rangle^2}\right\}
      \right]}
     {\exp\left[\frac{-1}{2(N-1)}\sum^{k}_{i=1}\left\{
       \sum_{\mu=1}^3\frac{p_{i,\mu}^2}{\langle p_\mu^2\rangle}
       +
       \frac{\left( E_i - \langle E\rangle\right)^2}{\langle E^2 \rangle - \langle E \rangle^2}\right\}\right]}  \nonumber
\end{eqnarray}

and  Equation~\ref{eq:JYOc2firstorderGeneral} becomes
\begin{align}
\label{eq:JYOc2firstorder}
& C(p_1,p_2)=1- \\
& \frac{1}{N}\left(
     2\frac{\vec{p}_{1,T}\cdot\vec{p}_{2,T}}{\langle p_T^2 \rangle} + 
      \frac{p_{1,z}\cdot p_{2,z}}{\langle p_z^2 \rangle} +
      \frac{\left(E_1-\langle E \rangle\right)\left(E_2-\langle E \rangle\right)}{\langle E^2\rangle-\langle E\rangle^2}\right) , \notag
\end{align}
where we have taken $\langle p_x^2 \rangle = \langle p_y^2 \rangle = \langle p^2_T \rangle/2$ in the azimuthally-symmetric case of interest.
In what follows and in Figures~\ref{fig:GenbodMult9LCMSK0p5}-\ref{fig:GenbodMult18-withEtaCut}, we shall refer to the first, second, and third terms within the parentheses of Equation~\ref{eq:JYOc2firstorder}
as the ``$p_T$'' ``$p_z$'' and ``$E$'' components, respectively.


If we somehow know $N$, $\langle p_T^2 \rangle$, $\langle p_z^2 \rangle$, $\langle E^2 \rangle$ and $\langle E \rangle$,
we can calculate EMCICs using Equation~\ref{eq:JYOc2}.  (See, however, the discussion at the start of the next Section.)
Better yet, if $N$ is large enough, then we can use Equation~\ref{eq:JYOc2firstorder}.
This is what is done in Figures~\ref{fig:GenbodMult9LCMSK0p5}-\ref{fig:GenbodMult18-withEtaCut}.
The open circles and orange inverted triangles represent the results of Equation~\ref{eq:JYOc2} and
Equation~\ref{eq:JYOc2firstorder}, respectively. 
The black circles, blue stars, and red triangles 
show the individual components of Equation~\ref{eq:JYOc2firstorder}; this decomposition will be 
relevant when we discuss the ``experimentalist's formula'' in the next Section.


Figures~\ref{fig:GenbodMult9LCMSK0p5}-\ref{fig:GenbodMult18-withEtaCut} make clear that each of the three terms in Equation~\ref{eq:JYOc2firstorder} produces non-trivial behavior of the \Alm's. Also clear is the importance of not neglecting the energy term. 
 We find also that the $p_z$ term affects $A_{2,2}$; this may be
surprising since $A_{2,2}$ quantifies the behavior of the correlation function in the ``out-side'' plane, while $\hat{z}$ is
the ``long'' direction in the Bertsch-Pratt system.
Clearly, EMCICs projected onto a 2-particle space are non-trivial objects.

The first-order expansion (Equation~\ref{eq:JYOc2firstorder}) agrees well
with the full expression (Equation~\ref{eq:JYOc2}) well for $N \gtrsim 10$.  Such multiplicities are relevant for $p+p$
measurements at RHIC (recalling that $N$ includes all particles, even unmeasured ones).  We see also that
the analytic calculations (open circles and inverted triangles) approximate the results of the \genbod simulation (green squares),
especially as the multiplicity and total energy of the event increases; increasing agreement for large $N$ and $E_{tot}$ is expected,
given the approximations leading to our analytic expressions.
We observe also that the analytically-calculated expressions respond identically to the kinematic cuts as does the simulation
(c.f. Figures~\ref{fig:GenbodMult18-noEtaCut} and ~\ref{fig:GenbodMult18-withEtaCut}).

Finally, the analytic calculations never reproduce {\it exactly} the simulations; we discuss this further in the next Section.

\section{AN EXPERIMENTALIST'S FORMULA}
\label{sec:formula}

Even for large $N$ and energy, the calculations do not exactly reproduce the EMCIC effects in the simulation.
One reason for this may be found, in fact, in the definition of the average values (e.g. $\langle p_z^2 \rangle$) themselves.
In Equation~\ref{eq:defineRMS}, average quantities are calculated using the distribution $\tilde{f}(p)$, which is
not affected by EMCICs.  Naturally, the only measurable distribution available to the experimentalist (even when \genbod
simulations serve as the ``experiment'') is $\tilde{f}_c(p)$.

Thus, it appears the experimentalist cannot plug her data into the equations~\ref{eq:defineRMS}
and~\ref{eq:JYOc2firstorder} to fully calculate EMCICs.  However, such an ambition would have been hopeless anyhow.
After all, even the total multiplicity $N$ (again, including photons etc) is rarely fully measured, and in principle $N$ is a number of ``primary'' particles, a murky concept in itself.  

To the practicing femtoscopist, there is a natural solution.
Having at hand (1) educated guesses for the quantities $N$, $\langle E^2 \rangle$ etc, and (2) a physically-motivated
functional form which connects these quantities to the correlations,  one may perform a fit.
Let us rewrite Equation~\ref{eq:JYOc2firstorder} as
\begin{eqnarray}
\label{eq:formula}
C\left(p_1,p_2\right) = 1
   - M_1\cdot\overline{\left\{\vec{p}_{1,T}\cdot\vec{p}_{2,T}\right\}}
   - M_2\cdot\overline{\left\{p_{1,z}\cdot p_{2,z}\right\}} \quad \\
   - M_3\cdot\overline{\left\{E_1 \cdot E_2\right\}}
   + M_4\cdot\overline{\left\{E_1 + E_2\right\}}
   - \frac{M_4^2}{M_3} , \qquad \nonumber
\end{eqnarray}
where
\begin{eqnarray}
\label{eq:fitParameters}
M_1\equiv\frac{2}{N\langle p_T^2 \rangle}
  \quad , \quad  M_2\equiv \frac{1}{N\langle p_z^2 \rangle} \qquad \qquad \qquad \nonumber \\
M_3\equiv\frac{1}{N\left(\langle E^2\rangle-\langle E\rangle^2\right)}  
  \quad , \quad M_4\equiv\frac{\langle E\rangle}{N\left(\langle E^2\rangle-\langle E\rangle^2\right)} .
\end{eqnarray}

The notation $\overline{\left\{X\right\}}$ in Equation~\ref{eq:formula} highlights 
the fact that $X$ is
a two-particle quantity which depends on $p_1$ and $p_2$ (or $\vec{q}$, etc).
From a practical point of view, $X$ will be averaged over the same $\vec{q}$ bins 
as used for the correlation function.  
For infinitesimally narrow $q$-bins,  $\overline{\left\{X\right\}}=X$.
The binned functions $\overline{\left\{X\right\}}$ then automatically 
reflect the same event and particle selection as the correlation function.
This involves nothing more than adding four more histograms to the several 
already being constructed by the experimentalist
as she processes pairs in the data.

Here, we should  emphasize that, in Equation~\ref{eq:formula}, $\vec{p_1}$, $E_1$, 
$\vec{p_2}$, and $E_2$ should be calculated in the collision center-of-momentum (CCM) frame. 
The reason is that Equation~\ref{eq:JYO1} (hence Eqs.~\ref{eq:CLT_FmGeneral}-\ref{eq:fitParameters}) 
assumes some fixed total energy and momentum to be conserved.
The event's total energy and momentum (hence $\langle E \rangle$, $\langle \vec{p} \rangle$, etc
appearing in Eqs.~\ref{eq:CLT_FmGeneral}-\ref{eq:fitParameters}) are fixed quantities in any given frame. 
In a {\it pair}-dependent frame (e.g. LCMS), the total energy and momentum of the event
will fluctuate, pair-by-pair.
Thus, while the correlation function may be {\it binned} in whatever frame one chooses,
the momenta $p^{\mu}_{i}$ on the right side of Eqs.~\ref{eq:JYO1}-\ref{eq:fitParameters} 
must be calculated in a pair-independent frame. In fact, starting with Equation~\ref{eq:comFrame}, 
we have chosen the CCM, for simplicity.

The parameters $M_i$ defined in Equation~\ref{eq:fitParameters}, on the other hand, are global and independent
of $p_1$ and $p_2$.  It is these which we will use as fit parameters.  The task is then fast and straightforward;
the EMCIC part of the correlation function $C(\vec{q})$ is simply a weighted sum of four functions.  Indeed, 
one may calculate coefficients as in Equation~\ref{eq:Alm} for the four new functions.  For example
\begin{equation}
A_{l,m}^{p_Z}\left(Q\right) \equiv
\sum_{{\rm bins} \thinspace i} \overline{\left\{p_{1,z}\cdot p_{2,z}\right\}}\left(Q,\cos\theta_i,\phi_i \right) \cdot
 Y_{l,m}\left(\cos\theta_i,\phi_i\right) ,
\end{equation}
etc.
Then, thanks to the linearity of Equation~\ref{eq:formula} and the orthonormality of $Y_{l,m}$'s, the measured \Alm's themselves are similarly
just weighted sums of harmonics
\begin{eqnarray}
\label{eq:AlmFit}
\AlmQ = \delta_{l,0}\cdot\left(1-M_4^2/M_3\right) 
  - M_1\cdot A_{l,m}^{p_T}\left(Q\right)             \qquad \\
  - M_2\cdot A_{l,m}^{p_Z}\left(Q\right)
  - M_3\cdot A_{l,m}^{(E\cdot E)}\left(Q\right)
  + M_4\cdot A_{l,m}^{(E+E)}\left(Q\right) . \nonumber
\end{eqnarray}
Treating Equation~\ref{eq:AlmFit} as a fit, we have a few (say six, for $l\leq 4$)
{\it one}-dimensional functions to fit with four adjustable weights.  The number
of degrees of freedom in this four-parameter fit remains high: $\sim 300$, for
six \Alm's, each with 50 bins in $|Q|$.

\begin{figure}[t!]
{\centerline{\includegraphics[width=0.5\textwidth]{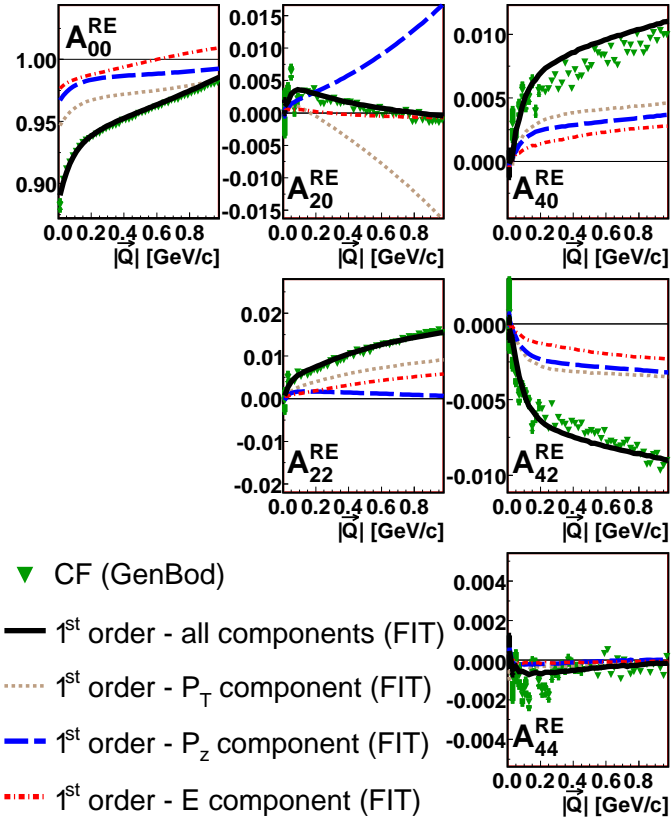}}}
\caption{(Color online) Green inverted triangles show the $\pi^--\pi^-$ correlation function, in the pair rest frame,
from 18-pion \genbod-generated events.
The black curve is a result of a fit with the ``experimentalist's formula'' from Equation~\ref{eq:AlmFit}.
Other curves represent the three component terms of the fit: 
$M_1\cdot A^{p_T}_{l,m}(Q)$ in the brown dotted line; 
$M_2\cdot A^{p_Z}_{l,m}(Q)$ in the blue dashed line;
$M_3\cdot A^{E\cdot E}_{l,m}(Q) + M_4\cdot A^{E+E}_{l,m}(Q)$ in the red dash-dot line.
  See text for details.\label{fig:FitAlmGenbod}}
\end{figure}

An example is shown in Figure~\ref{fig:FitAlmGenbod}, where \genbod-calculated correlation
functions are fitted with the form of Equation~\ref{eq:AlmFit}.  Not surprisingly, the minimization
procedure returned fit parameters $M_i$ very close to the values calculated via Equation~\ref{eq:fitParameters}.
However, exact agreement between the ``best'' parameter values returned by the fit, and those from
Equation~\ref{eq:fitParameters} is not expected.  This is because the large-$N$ approximation is only approximately
valid and due to the fact that $\tilde{f}(p)\neq\tilde{f}_c(p)$, as discussed previously.
Treating the $M_i$
as adjustable parameters leads to a slightly different weighting of the terms, and a slightly better fit to the data.

Our original goal was not to understand EMCICs {\it per se}, but to extract the femtoscopic 
information from measured two-particle correlations.
Assuming that the only non-femtoscopic correlations are EMCICs, one may simply add the
femtoscopic terms $\Phi_{\rm femto}(p_1,p_2)$ to the fitting function
in Equation~\ref{eq:formula} or~\ref{eq:AlmFit}.

\begin{eqnarray}
\label{eq:FemtoFitWithEmcics}
C\left(p_1,p_2\right)&=& \Phi_{\rm femto}(p_1,p_2) \times  \\
   &\Big(&1 - M_1\cdot\overline{\left\{\vec{p}_{1,T}\cdot\vec{p}_{2,T}\right\}}
   - M_2\cdot\overline{\left\{p_{1,z}\cdot p_{2,z}\right\}} \nonumber \\
   &-& M_3\cdot\overline{\left\{E_1 \cdot E_2\right\}}
   + M_4\cdot\overline{\left\{E_1 + E_2\right\}}
   - \frac{M_4^2}{M_3}\Big) . \nonumber  
\end{eqnarray}

Common femtoscopic fitting functions (e.g. Gaussian in the out-side-long space) usually contain 
$\sim 5$ parameters (e.g. $N$, $\lambda$, $R_{i}$, $i=o,s,l$).
In the imaging technique~\cite{Brown:1997ku},
one assumes the separation distribution is described by a sum of splines, rather than a
Gaussian; here, too, there are usually
4-5 fit parameters (spline weights).
So, by including EMCIC effects, we have roughly doubled the number of fit parameters, relative to
a ``traditional'' fit which ignores them.
This is a non-trivial increase in analysis complexity.  However, we keep in mind two points.

Firstly, the increased effort is simply necessary.  EMCICs (and possibly other
important non-femtoscopic correlations) are present and increasingly relevant at low multiplicity.
One option is to ignore them, as has sometimes been done in early high-energy experiments.  However, 
with the new high-quality data and desire for detailed understanding at RHIC,
ignoring obvious features such as those presented in~\cite{Chajecki:2005zw} is clearly unacceptable.
Perhaps a slightly better option is to invent an ad-hoc functional form~\cite{Agababyan:1996rg}
without a strong real physical basis.
We hope that the results here
present a relatively painless and more reasonable, third option.

Secondly, while the non-femtoscopic EMCICs are not confined to the large-$Q$ region (an important point!), the femtoscopic correlations are confined
to the small-$Q$ region.  Therefore, one hopes that the addition of four new parameters to the fit of the correlation function will not render the
fit overly unwieldy.  While we cannot expect complete block-diagonalization of the fit covariance matrix, one hopes that the $M_i$ are determined
well enough at high $Q$ that the femtoscopic fit parameters can be extracted at low $Q$.

\section{Non-identical particle correlations}
\label{sec:nonid}

For at least two reasons, it is important to turn attention to correlations between non-identical particles.

First, it is natural to ask whether one can use other particle combinations to ``correct'' for effects of EMCICs in,
say, identical-pion correlation functions.
After all, EMCICs are induced by global constraints on the entire event, not a specific particle species.
For example, various experiments have explored using  $(\pi^+, \pi^-)$ correlations to account for EMCICs
in  $(\pi^+, \pi^+)$ correlation functions~\cite{Avery:1985qb,collaboration:2007he,Chekanov:2003gf}.

Second, it is also important to know whether EMCICs could cloud the interpretation of correlations between non-identical
particles.  It is increasingly common to study asymmetries in the correlation functions of, say $\pi-K$ pairs~\cite{Adams:2003qa},
interpreting such as a ``shift'' in the average point of emission between the two particles~\cite{Lednicky:1995vk}.
In the spherical harmonic decomposition, such shifts appear in the $l=1$ moments (c.f. Appendix~\ref{app:symmetries}).
We will find that EMCICs can indeed generate an asymmetry which might naively be considered proof of a femtoscopic shift.

Here we discuss two effects-- one immediately obvious and one more subtle-- which
are relevant for the above issues.  The discussion is broken into three parts.  Neglecting EMCICs and any other source of correlation 
at first, we briefly show the effects of two common resonances on correlations between oppositely-charged pions
in a toy model.  Thus calibrated, we use the more realistic and complex \pythia model to illustrate a non-trivial
interplay between EMCICs and the resonances, which can mock up a femtoscopic asymmetry signal.  Finally, we return
to a toy model-- now with non-identical particles and EMCICs, but without resonances or the several other sources of correlation
present in \pythia-- to make clear the mechanism behind the special effects EMCICs have on non-identical particle correlations.

\subsection{Effect of resonances}
\label{sec:toymodel}
First we consider the effect of resonances. To focus on effects other than global EMCICs we use a toy model in which only ten identical resonances per event are generated and no other particles. The momentum of each resonance is generated from a thermal  distribution;  energy and momentum are conserved for each decay separately, but not globally for the whole event.

\begin{figure}[t]
{\centerline{\includegraphics[width=0.5\textwidth]{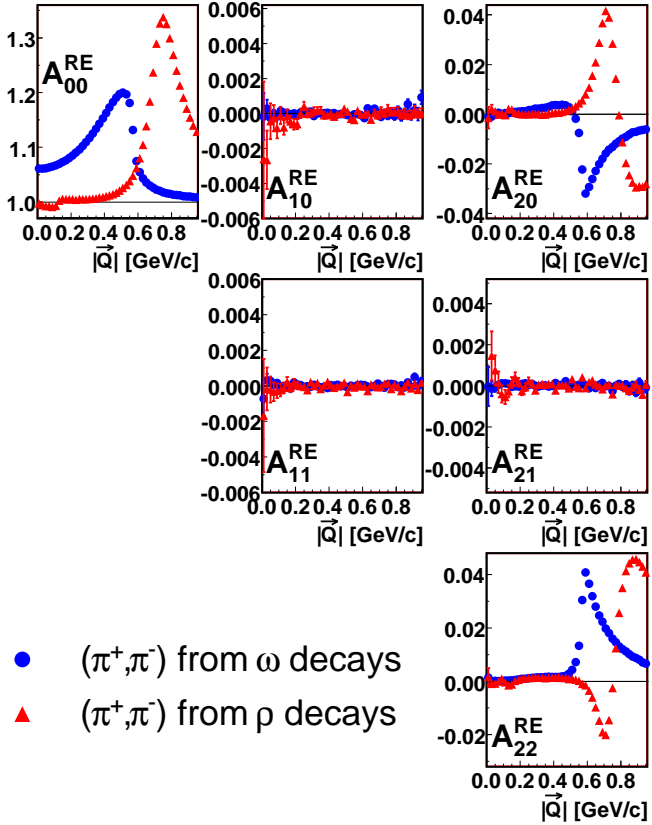}}}
\caption{(Color online) $(\pi^{+},\pi^{-})$ correlation functions calculated in the LCMS frame for events including $\omega$ (blue squares) and $\rho$ (red triangles) resonances . \label{fig:omegarhoLCMS}}
\end{figure}

Figure~\ref{fig:omegarhoLCMS} shows the spherical harmonic moments of $(\pi^{+},\pi^{-})$ correlation functions for events including $\omega$ (blue squares)  and $\rho$ resonances (red triangles).
As seen, even without considering EMCICs, the correlations among particles coming from resonance decays produce non-trivial structure. In this case,  one cannot simply divide $(\pi^{+},\pi^{+})$ correlation function by $(\pi^{+},\pi^{-})$ to ``remove'' EMCICs.

\begin{figure}[t]
{\centerline{\includegraphics[width=0.49\textwidth]{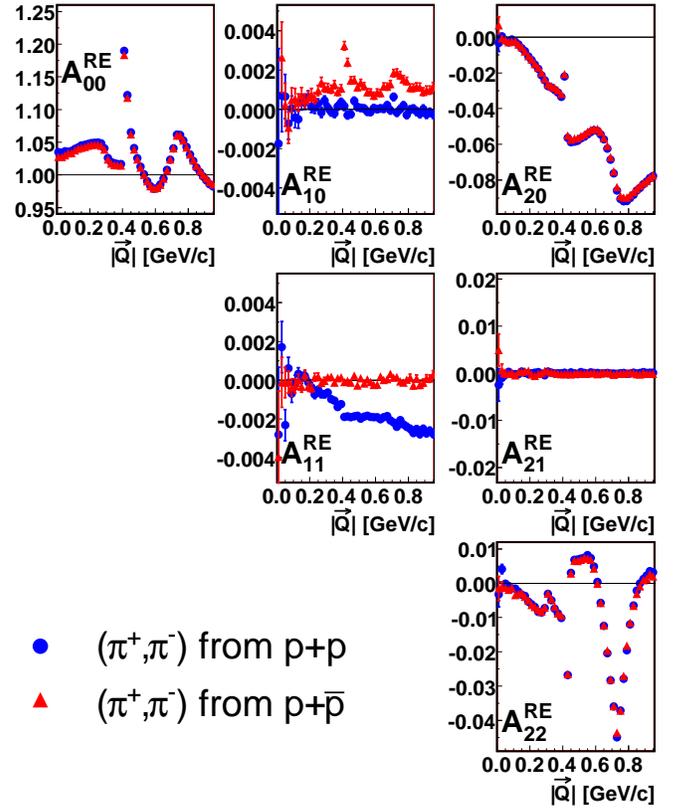}}}
\caption{(Color online) SHD moments of $(\pi^{+},\pi^{-})$ correlation function from $p+p$ and $p+\bar{p}$ collisions at 200~GeV calculated from \pythia events. \label{fig:pythiaNonId}}
\end{figure}

\subsection{Entrance channel asymmetries}
\label{sec:entranceChannel}
In addition to the correlation between daughters of resonance decays (c.f. Fig.~\ref{fig:omegarhoLCMS}), there is a more 
subtle effect to consider. This happens when the two particles have different inclusive momentum distributions 
{\it and} energy and momentum are globally conserved.  Under these conditions non-identical particle  correlations exhibit 
structure absent in identical particle correlations.

Figure~\ref{fig:pythiaNonId} shows \pythia~\cite{Sjostrand:2000wi} calculations of $(\pi^{+},\pi^{-})$ correlations for 
$p+p$ and $p+\bar{p}$ collisions at 200 GeV.
In addition to obvious correlations between daughters of resonance decays ($K_{s}^{0}$, $\omega$, $\rho$), we see 
additional structure.  We focus on the structure in the $l=1$ moments.
In general, such moments
need not vanish for correlations of non-identical particles,
as discussed in Appendix~\ref{app:symmetries}.

Correlations between sibling daughters of $\rho$ and $\omega$ resonance decay do not generate $l=1$ moments,
as seen in Section~\ref{sec:toymodel}.  However, pions which are daughters of these decays will in general have a different
single-particle momentum distribution than pions from other sources in the event.  If the fraction of pions from resonance
decay, as a function of pion momentum, is different for $\pi^+$ and $\pi^-$, then the single-particle distributions of
positive and negative pions will be different.  We argue below that it is this difference
in single-particle distributions which is the key to the non-vanishing $l=1$ moments; that this difference may arise from
resonances in the case at hand is irrelevant.

In the $p+\bar{p}$ collisions, the fraction of $\pi^+$ coming from any given source (e.g. $\rho$-decay) must be identical to that
of $\pi^-$, for a given value of $p_T$.  Thus the $p_T$ distribution of $\pi^-$ must be identical to that of $\pi^+$.  However, the rapidity
distributions will be mirror images of each other.  Thus, any asymmetry in $\pi^--\pi^+$ correlations from $p+\bar{p}$ collisions
will be associated with $q_{\rm long}$, and will appear in $A_{1,0}$, as
seen in Figure~\ref{fig:pythiaNonId}.  Similarly, the vanishing (non-vanishing) moment $A_{1,0}$ ($A_{1,1}$) for $p+p$ collisions
reflects the fact the single-particle distributions will show no asymmetry in rapidity, but may differ as a function of
$p_T$.

Since single-particle distributions divide out of a correlation function, a difference between $\pi^+$ and $\pi^-$ momentum distributions,
by itself, cannot generate a signal in \Alm's.  Rather, a global correlation, coupled with this difference, generates the signal.
We discuss this further below.

\subsection{A simpler case}

In Section~\ref{sec:entranceChannel}, we argued that the small difference in single-particle momentum
distributions between positive and negative pions produced by \pythia, coupled with global conservation laws,
generated non-trivial EMCICs in the non-identical particle correlations.  However, \pythia contains many
non-EMCIC sources of correlations, related to string fragmentation and other processes, which might be
flavor/isospin-dependent.  To make clearer our argument, we here show a simple \genbod simulation, containing
both pions and protons, but no explicit correlations between them such as a $\Delta$ resonance.
Due at least to their different masses, $\tilde{f}_{\rm proton}\neq\tilde{f}_{\rm pion}$ is guaranteed.

\begin{figure}[t]
{\centerline{\includegraphics[width=0.49\textwidth]{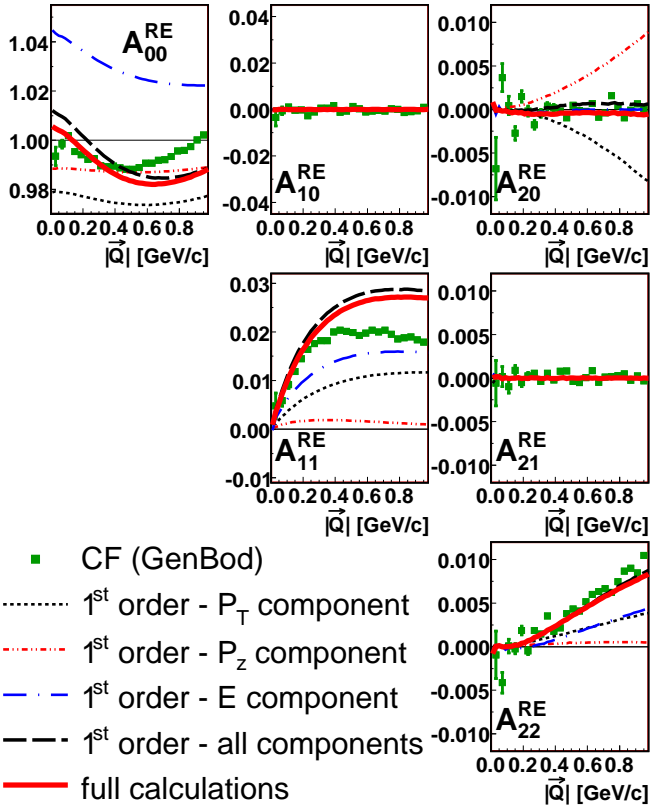}}}
\caption{(Color online) 
Green squares show the $\pi-p$ correlation function, in the pair rest frame,
from \genbod-generated events.
Gold inverted triangles are the SHD coefficients of Equation~\ref{eq:JYOc2} for k=2.
Black circles, blue stars and red triangles are SHD coefficients of the first, second and third terms, respectively, of the
right side of Equation~\ref{eq:JYOc2firstorder}.  Open circles are SHD coefficients of the right side of Equation~\ref{eq:JYOc2firstorder}.
\label{fig:genbodNonIdPairCMS}}
\end{figure}

Figures~\ref{fig:genbodNonIdPairCMS} shows the $\pi-p$ correlation function.  Since the underlying single-particle
proton and pion distributions are isotropic, $A_{1,0}$ (sensitive to shape elongation in $C(\vec{q})$ in \ql relative
to transverse components) is expected to vanish.  $A_{1,1}$ is finite, however, due to differences in $p_T$ distributions.
Since there is no other source of correlation in the simulation, this obviously is an EMCIC.

From Figure~\ref{fig:genbodNonIdPairCMS} it is also clear that neither Equation~\ref{eq:JYOc2} nor its first-order
expansion~\ref{eq:JYOc2firstorder} fully describes the correlation function.  This is due to the fact that our formalism
is built on the assumption that all particles in the system follow the same parent distribution, as pointed out
after Equation~\ref{eq:defineRMS}.

\section{Summary}
\label{sec:summary}

To truly claim an understanding of the bulk nature of matter at RHIC and the LHC, a detailed picture of the
dynamically-generated geometric substructure of the system created in heavy ion collisions
is needed.  It is believed that this substructure,
and the matter itself, is dominated by strong collective flow.  The most direct measure of this flow is
a measurement of the space-momentum correlation (e.g. $R\left(m_T\right)$) it generates.  The physics
of this large system, and the signals it generates, should be compared to the physics dominating $p+p$
collisions, as is increasingly important in high-$p_T$ studies at RHIC.  
For small systems, however,
non-femtoscopic effects contribute significantly to the correlation function, clouding the extraction
and interpretation of femtoscopic ones.  

We have discussed a spherical harmonic representation of the correlation function which clearly separates
components of the three-dimensional shape measured in modern experiments.  This representation is maximally
efficient, inasmuch as only a few one-dimensional plots need be examined to extract full 3-D shape information.
The relevant number of such plots is further reduced due to symmetry conditions, discussed in detail in
Appendix~\ref{app:symmetries}.

EMCICs, correlations generated by kinematic conservation laws, are surely present and increasingly relevant
as the event multiplicity is reduced.  Using the code \genbod to study correlation functions solely driven
by EMCICs, we found highly non-trivial three-dimensional structures strongly influenced by event characteristics (multiplicity
and energy) and kinematic particle selection.

We extended the work of Danielewicz, Ollitrault and Borghini to include
four-momentum conservation and applied it to correlation functions commonly used in femtoscopy.  We found
structures associated individually with the conservation of the four-momentum components, which interfere
in nontrivial ways.  
Comparison of the analytic EMCIC calculations with the \genbod simulation gave confidence that
the approximations (e.g. ``large'' multiplicity $N$) entering into the calculation were sufficiently
valid, at least for multiplicities considered here.
We further showed that the full EMCIC calculation can safely be replaced
with a first-order expansion in $1/N$.

Based on this first-order expansion, we developed a practical,
straight-forward ``experimentalist's formula''
to generate histograms from the data which are later used in a generalized fit to the
measured correlation function, including EMCICs and femtoscopic correlations.
The degree to which this functional form fully describes measured experimental
correlation functions has not been discussed and will need exploration on a case-by-case
basis.

There is strong interest in correlations between non-identical particles, for two reasons.
Firstly, sometimes $\pi^--\pi^-$ correlations are divided by $\pi^+-\pi^-$ correlations in
an attempt to ``divide out'' EMCICs.  (In such a procedure, resonance regions are avoided, naturally.)
We discussed potential problems with such an approach, related to entrance-channel asymmetries coupled
with EMCICs.  Secondly, 3-D asymmetries (sometimes quantified as ``double ratios'', e.g.~\cite{Adams:2003qa})  in the correlation
function for different-mass particles (e.g. $\pi-K$) are often interpreted in terms of dynamically-generated
differences in the average space-time emission point between the two particles.  Using a very simple
example, we discussed that EMCICs might significantly cloud such an interpretation.

The huge systematics of results and interest in femtoscopy in heavy ion collisions is
renewing similar interest in the space-time signals from $p+p$ collisions.  Direct comparisons
between the two systems are now possible at RHIC and have already produced intriguing  preliminary
results.  Very soon, $p+p$ collisions will be measured in the LHC experiments, and the heavy ion experimentalists
will be eager to apply their tools.  The femtoscopic tool is one of the best in the box-- so long as
we keep it sufficiently calibrated with respect to non-femtoscopic effects increasingly relevant in small systems.

\medskip

\begin{acknowledgments}
We wish to thank Drs. Mark Baker,
Nicolas Borghini, Pawe\l~Danielewicz, Ulrich Heinz, Adam Kisiel, Konstantin Mikhaylov, Dariusz Mi\'{s}kowiec, 
Jean-Yves Ollitrault, Alexey Stavinsky, and Boris Tom\'{a}\v{s}ik for important suggestions and insightful discussions.
We would also like to thank an anonymous referee for useful suggestions to Section~\ref{sec:JYO}. 
This work supported by the U.S. National Science Foundation under Grant No. PHY-0653432.

\end{acknowledgments}

\vspace*{1cm}

\appendix

\section{EMCIC effects on $v_2$}
\label{app:V2}

\begin{figure}[t!]
{\centerline{\includegraphics[width=0.5\textwidth]{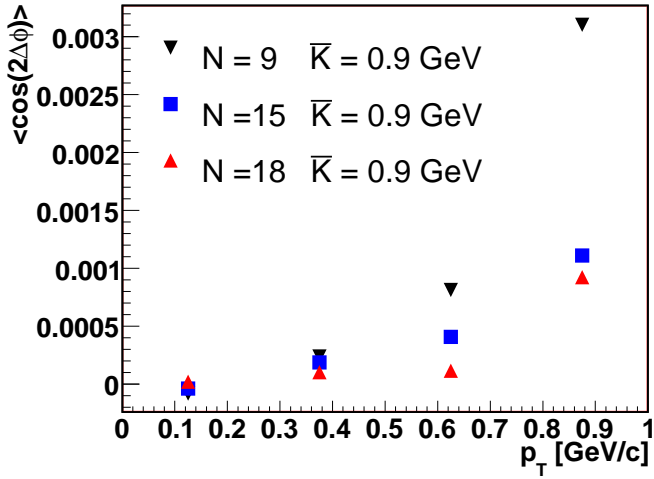}}}
 \caption{(Color online) $v_2(p_T)$ for different event multiplicities. See text for details. \label{fig:v2vspT}}
\end{figure}

Since EMCICs can produce a structure in the correlation function even in the absence of femtoscopic
correlations, it is worthwhile to check analytically and then confirm with simulations whether $v_2$-- a
common measure of collective elliptic flow~\cite{Ollitrault:1992bk}-- may be affected by EMCICs.

When calculating flow from two-particle correlations 
we have the following relations 
\begin{eqnarray}
\label{eq:V2}
\int \cos\big(m \Delta \phi\big) \cos\big(n \Delta \phi\big) d\Delta\phi = \delta_{m,n}\pi,
\end{eqnarray}
where for $v_2$, $n=2$.

This means that in the absence of flow all EMCIC terms vanish  
except for the ones that exhibit $\cos\big(2\Delta\phi\big)$ 
dependence of $\Delta\phi$. For example in the first order expansion 
of EMCICs (see Eq.~\ref{eq:JYOc2}) 
there is a term $\vec{p}_{T,1}\vec{p}_{T,1}\sim \cos\big(\Delta\phi\big)$. 
This term gives no contribution to $v_2$, nor do any other terms from $1/N$
expansion. The first term that gives a non-zero contribution to $v_2$  
(means, goes like $\cos\big(2 \Delta \phi\big)$) is the second order 
expansion term in $\vec{p}_{T}$ that is proportional to 
$\big(\vec{p}_{T,1}\vec{p}_{T,1}\big)^{2} \sim \cos^2\big(\Delta\phi\big) \sim \cos\big(2\Delta\phi\big)$.
This term (as well as a few other terms in higher-order $1/N$ expansion) 
will give a non-zero contribution to $v_2$.
In our \genbod simulations we do not have a flow so we can study 
the magnitude of the EMCIC effects on $v_2$ measurements. 
Such results are presented on Figure~\ref{fig:v2vspT} where we plot 
$v_2$ vs $p_T$ for three different event multiplicities while 
the free kinetic energy per particle is fixed ($\bar K=0.9~GeV$).

As seen, the magnitude of a non-flow contribution to $v_2$ from EMCICs 
is getting smaller with increasing multiplicity and even 
for low-multiplicity events the magnitude is of order 
of a few per-mile for large $p_T$. 
From this dependence we can predict that this effect will be so small 
in heavy ion collisions that it can be simply neglected.

\section{Symmetry Considerations}
\label{app:symmetries}

The spherical harmonic decomposition representation, in which three-dimensional correlation functions
are represented by several one-dimensional moments, \Alm, efficiently condenses the shape information.
A much {\it greater} increase in efficiency comes, however, with the realization that many \Alm's must
vanish by symmetry, depending on the cuts and conditions of the analysis.  Besides reducing information
by significant factors, this realization also provides diagnostic power-- non-physical artifacts often
appear in \Alm's which do not vanish when they should.  Digging out such effects in the traditional
three-dimensional Cartesian representation can be quite difficult.

\renewcommand{\theenumi}{[\Alph{enumi}]}

In the most general case, the 3-D correlation function may have any
shape, with no symmetry constraints.  In this case, none of the \Alm's need vanish.
Usually, however, an analysis is less than fully general, and symmetry consequences then arise.

In particular, we will consider four common conditions used in practice:
\begin{enumerate}
\item One measures correlations between identical particles               \label{cut:ID}         
\item The measurement covers a symmetric rapidity region about y=0 
      and the collision is between identical ions (e.g. Au+Au
      rather than Au+Si)                                                  \label{cut:Midrapidity}  \label{cut:y=0}
\item The measurement is integrated over reaction-plane angle             \label{cut:RPintegrated} \label{cut:noRP}
\item The measurement might be correlated with the {\it second}-order
      reaction-plane, but the first-order reaction-plane is not known.    
      In other words, the direction of the impact parameter is known at best
      only modulo $\pi$.                                                  \label{cut:RP2only}      \label{cut:RP2}
\end{enumerate}

Our strategy begins by identifying transformations in relative momentum $\vec{q}$ under which
the measured correlation must be invariant.  As an example, since the overall sign of $\vec{q}$
is meaningless when discussing pairs of identical particles (condition~\ref{cut:ID}), 
$C\qcomp{}{}{} = C\qcomp{-}{-}{-}$, or, in spherical coordinates, $C(Q,\cos\theta,\phi)=C(Q,-\cos\theta,\phi-\pi)$.

We then use a symmetry of the spherical harmonics, here $\YlmArg{}{}{}=(-1)^{l}\YlmArg{-}{}{+\pi}$
to find
\begin{align}
&\Alm(Q)  \equiv  \frac{1}{4\pi} \int^{2\pi}_{0}   d\phi \int^{1}_{-1} d\cos\theta C(Q,\cos\theta,\phi)\YlmArg{}{}{}     \nonumber \\
&         =       \frac{1}{4\pi} \int^{2\pi}_{0}   d\phi \int^{1}_{-1} d\cos\theta C(Q,-\cos\theta,\phi-\pi)\YlmArg{}{}{}   \nonumber \\
&         =       \frac{1}{4\pi} \int^{\pi}_{-\pi} d\phi \int^{-1}_{1} (-d\cos\theta) C(Q,\cos\theta,\phi)\YlmArg{-}{}{+\pi}   \nonumber \\
&         =       \frac{1}{4\pi} \int^{2\pi}_{0}   d\phi \int^{1}_{-1} d\cos\theta C(Q,\cos\theta,\phi)(-1)^l\YlmArg{}{}{}) \nonumber \\
&         =       (-1)^{l}\Alm(Q) .
\end{align}
Thus, all odd-$l$ moments \Alm must vanish, for correlations between identical particles.

The same type of reasoning is used below, in identifying symmetry constraints for various combinations
of analysis conditions.

\subsection{$\vec{q}$ transformations and \Ylm response}

Table~\ref{tab:qTransformations} lists all combinations in which one or more of the components of $\vec{q}$ can
change sign.  For later reference, the transformations are numbered 0\dots 7, according to a binary scheme.
The effect of the transformation on the spherical harmonics appears in the last column of the Table.

\begin{table}
\begin{tabular}{|l|l|l|}
\hline
\# &  transformation     & \Ylm consequence \\
\hline
0  & \transform{+}{+}{+} & $\Ylm\rightarrow\Ylm$ \\
\hline
1  & \transform{+}{+}{-} & $\Ylm\rightarrow\prty{l+m}\Ylm$ \\
\hline
2  & \transform{+}{-}{+} & $\Ylm\rightarrow\Ylm^*$ \\
\hline
3  & \transform{+}{-}{-} & $\Ylm\rightarrow\prty{l+m}\Ylm^*$ \\
\hline
4  & \transform{-}{+}{+} & $\Ylm\rightarrow\prty{m}\Ylm^*$ \\
\hline
5  & \transform{-}{+}{-} & $\Ylm\rightarrow\prty{l}\Ylm^*$ \\
\hline
6  & \transform{-}{-}{+} & $\Ylm\rightarrow\prty{m}\Ylm$ \\
\hline
7  & \transform{-}{-}{-} & $\Ylm\rightarrow\prty{l}\Ylm$ \\
\hline
\end{tabular}
\caption{The possible transformations (numbered in the left column) in which the signs of $\vec{q}$ components flip, and
the effect of the transformation on the \Ylm's.
\label{tab:qTransformations}}
\end{table}

Transformation (0), of course, is the trivial identity transformation, under which any
correlation function is invariant, and which imposes no symmetry constraint.  We include it in the
Table only for completeness, and do not discuss it further.

\begin{table}[ht]
\begin{tabular}{|c|c|c|c||c|c|c|c|c|c|c||l|}
\hline
\multicolumn{4}{|c||}{Anal. Conditions} & \multicolumn{7}{|c||}{$C(\vec{q})$ invariances} & which \Alm 's vanish  \\ 
\hline
\ref{cut:ID} & \ref{cut:y=0} & \ref{cut:noRP} & \ref{cut:RP2} &   1   &   2   &   3   &   4   &   5   &   6   &   7   & ~ \\
\hline
  \Surd      &       ~       &       ~        &       ~       &   ~   &   ~   &   ~   &   ~   &   ~   &   ~   & \Surd & $l$ odd  \\
\hline
  \Surd      &    \Surd      &       ~        &       ~       &   ~   &   ~   &   ~   &   ~   &   ~   &   ~   & \Surd & $l$ odd  \\
\hline
  \Surd      &      ~        &    \Surd       &   (\Surd)     &   ~   & \Surd &   ~   &   ~   & \Surd &   ~   & \Surd & \ReAlm : $l$ odd  \\
             &               &                &               &       &       &       &       &       &       &       & \ImAlm : $\forall l,m$ \\
\hline
  \Surd      &    \Surd      &    \Surd       &   (\Surd)     & \Surd & \Surd & \Surd & \Surd & \Surd & \Surd & \Surd & \ReAlm : $l$ and/or $m$ odd \\
             &               &                &               &       &       &       &       &       &       &       & \ImAlm : $\forall l,m$ \\
\hline
  \Surd      &      ~        &       ~        &    \Surd      &   ~   &   ~   &   ~   &   ~   &   ~   &   ~   & \Surd & $l$ odd  \\
\hline
  \Surd      &   \Surd       &       ~        &    \Surd      & \Surd &   ~   &   ~   &   ~   &   ~   & \Surd & \Surd & $l$ and/or $m$ odd \\
\hline
\hline
    ~        &       ~       &       ~        &       ~       &   ~   &   ~   &   ~   &   ~   &   ~   &   ~   &   ~   & ~~~~-~~~~ \\
\hline
    ~        &    \Surd      &       ~        &       ~       &   ~   &   ~   &   ~   &   ~   &   ~   &   ~   &   ~   & ~~~~-~~~~ \\
\hline
    ~        &      ~        &    \Surd       &   (\Surd)     &   ~   & \Surd &   ~   &   ~   &   ~   &   ~   &   ~   & \ImAlm : $\forall l,m$ \\
\hline
    ~        &    \Surd      &    \Surd       &   (\Surd)     & \Surd & \Surd & \Surd &   ~   &   ~   &   ~   &   ~   & \ReAlm : odd $(l+m)$  \\
             &               &                &               &       &       &       &       &       &       &       & \ImAlm : $\forall l,m$ \\
\hline
    ~        &      ~        &       ~        &    \Surd      &   ~   &   ~   &   ~   &   ~   &   ~   &   ~   &   ~   & ~~~~-~~~~\\
\hline
    ~        &   \Surd       &       ~        &    \Surd      & \Surd &   ~   &   ~   &   ~   &   ~   &   ~   &   ~   & odd $(l+m)$ \\
\hline

\end{tabular}
\caption{Symmetry consequences of analysis conditions.
The left four columns show various combinations of analysis cuts and conditions, identified [A]-[D] as discussed in
the beginning of this Appendix.  
(Note that condition~\ref{cut:noRP} implies condition~\ref{cut:RP2}; this is indicated by the symbol
$(\Surd)$ in column~\ref{cut:RP2}.)
The middle seven columns indicate the consequent invariance symmetries of the correlation function
according to the numbering scheme of Table~\ref{tab:qTransformations}.  The right-most column indicates which, if any, spherical
harmonic moments of the correlation function must vanish.
\label{tab:CFsymmetries}}
\end{table}

\subsection{Restrictions, invariants, and consequences on \Alm's}

Under which of the transformations in Table~\ref{tab:qTransformations} does the correlation function remain invariant?
Since identical-particle correlations are more common than correlations between non-identical particles, there will be
a greater familiarity with the symmetries of the former.  Thus, we begin with this more familiar case and then discuss
non-identical particle correlations.

\subsubsection{Correlations between identical particles}

To systematically identify those transformations in Table~\ref{tab:qTransformations} which leave a correlation function
invariant, it helps to have a concrete functional form to discuss.
For identical pions, the correlation function is often
parameterized as a Gaussian with six ``radius'' parameters,
\begin{align}
\label{eq:IdGaussian}
C\qcomp{}{}{} = 1+ \lambda\cdot\exp\left(-\RoS\qoS-\RsS\qsS-\RlS\qlS-\right.  \\ 
 \left. -2\RosS\qo\qs-2\RolS\qo\ql-2\RslS\qs\ql\right) . \notag
\end{align}
While measured correlation functions often have non-Gaussian features not captured by this parameterization,
the form given in Equation~\ref{eq:IdGaussian} contains the generic and most general symmetries of all correlation functions
using identical particles.  Thus, we use this familiar example to focus the discussion.
The six parameters of form~\ref{eq:IdGaussian} describe an ellipsoid described by three axis lengths, and rotated
by three Euler angles in $\vec{q}$-space.  Measured examples are shown and discussed in~\cite{Lisa:2000xj}.

Clearly, the form of Equation~\ref{eq:IdGaussian} is invariant under transformation (7),
as discussed earlier.  Invariance under any transformations (1)-(6) requires that one or
more of the ``radius'' parameters $R_{ij}^2$ vanish.  In general, none of them do~\cite{Lisa:2000xj,Lisa:2000ip},
even when considering a region symmetric about midrapidity in a collision between identical ions 
(condition~\ref{cut:Midrapidity})~\footnote{\label{fn:midrapiditySurprise}At first, it seems surprising that,
in the absence of reaction-plane assumptions, no additional symmetry
constraint is imposed onto the correlation function by a symmetric selection about midrapidity-- i.e. none
of the ``radius'' parameters $R_{ij}^2$ are required to vanish.  However, the selection does impose
symmetry constraints at a ``higher'' level.  In identical-particle correlations, for example, while $R^2_{ol}$ need not vanish at midrapidity
for any given measured correlation function, symmetry demands a relationship between $R^2_{ol}$ measured
in different correlation functions; in particular
$R^2_{ol}(\phi_{K,{\rm RP}}+\pi)=-R^2_{ol}(\phi_{K,{\rm RP}})$,
where $\phi_{K,{\rm RP}}$ is the angle between the total pair momentum and the reaction plane.
Symmetries at this level are discussed in detail in~\cite{Heinz:2002au}.}.

If the measurement is integrated over reaction plane angle, then the ``side'' direction has no relevant sign,
$\RosS=\RslS=0$,
and the correlation function is invariant under transformation (2).  While \RolS need not vanish~\cite{Chapman:1994yv}, the correlation
function is unchanged if \qo and \ql change sign together (transformation (5)).

Further constraining the measurement
to a symmetric region about midrapidity implies also that $\RolS$ vanish,
and the correlation function is then invariant under all transformations (0)-(7).
This is the most common set of measurement conditions.

At high energies, it is common only to determine the second-order reaction plane.  This corresponds to condition~\ref{cut:RP2}.
If the measurement is performed at midrapidity (condition~\ref{cut:y=0}), then \RosS is the only non-vanishing
cross-term radius, so the correlation function is invariant under transformation (1).
Away from midrapidity, \RolS need not vanish, so (7) is again the only remaining transformation leaving $C(\vec{q})$
invariant.

\subsubsection{Correlations between non-identical particles}

Correlations between non-identical particles are no longer invariant under transformation (7), as they
may depend on odd-power terms of the components of $\vec{q}$.  In the case of femtoscopic correlations,
the strengths of these odd powers probe asymmetries in the average emission point between the two particle
species~\cite{Lednicky:1995vk}.

From a symmetry standpoint, the correlation function will be characterized by nine parameters,
rather than the six ``HBT radii'' of Equation~\ref{eq:IdGaussian}.  In the simple case that $C\qcomp{}{}{}$ would
be Gaussian, these new parameters might represent the offset from the origin of the ellipsoid in $\vec{q}$-space.

In the absence of any cuts-- or if only the midrapidity condition~\ref{cut:Midrapidity} is applied~[\ref{fn:midrapiditySurprise}],
all nine parameters may take any value, and there are no required invariances or symmetry constraints.  If the reaction-plane
is integrated over (condition~\ref{cut:noRP}) then $C\qcomp{}{}{}$ may remain sensitive to the sign of \qo~(reflecting, for example, a
different average time of emission between the particles~\cite{Lednicky:1995vk}) and \ql~(reflecting the difference in emission
point in the beam direction, for analyses away from midrapidity), but not \qs, since an angle-averaged physical source must be symmetric
with respect to the beam axis.

Unlike the case in which it is the sole condition,
if the midrapidity condition~\ref{cut:Midrapidity} is imposed {\it together} with condition~\ref{cut:noRP}, then it does
have an effect.  In particular, a dependence on the sign of \ql vanishes.

If condition~\ref{cut:noRP} is relaxed to condition~\ref{cut:RP2} (i.e. the analysis is sensitive to the second-order reaction plane),
then the sign of \qs~ may matter.  This is because the sign of \qo~always affects correlations between non-identical particles and,
as in identical particle correlations in which \RosS~may be finite, so the sign of $\qo\qs$ may separately matter.  Thus, imposition of
\ref{cut:RP2} alone implies no symmetry constraints.

\newcommand {\ct} {\cos\theta}
\newcommand{\cti}{\cos\theta_i}
\newcommand{\Dct}{\Delta_{\cos\theta}}
\newcommand{\Dp}{\Delta_\phi}

\newcommand{\lpr}{l^\prime}
\newcommand{\mpr}{m^\prime}

\section{Finite binning effects}
\label{app:binning}

Equation~\ref{eq:Alm} defines the harmonic moments in terms of a continuous correlation function.
Most experimentally-measured correlation functions are constructed via histograms with discrete,
finite bins.  For decomposition into spherical harmonics, a natural choice would be to use bins in $Q$, $\ct$ and $\phi$ (c.f. Eq.~\ref{eq:SHDcoords}).
Here, we will find an approximate expression, analogous to Equation~\ref{eq:Alm}, for the harmonic moments in terms
of the discretized correlation function.

We denote the fixed bin sizes in the angular coordinates as $\Dct$ and $\Dp$.  Binning in $Q$ is unimportant here,
since $Q$ is carried as an explicit argument in both $C$ and \Alm.  The binned correlation function (denoted
with superscript $\Delta$) is related to the continuous one as
\begin{align}
\label{eq:binning}
&C^\Delta \left(Q,\cti,\phi_i\right) \nonumber \\
& = \frac{1}{\Dp\Dct}  
    \int_{\phi_i-\Delta_\phi/2}^{\phi_i+\Delta_\phi/2}d\phi 
    \int_{\cos\theta_i-\Delta\cos\theta/2}^{\cos\theta_i+\Delta\cos\theta/2}d(\cos\theta)
    C\left(Q,\ct,\phi\right) \nonumber \\
& = 
    \frac{\sqrt{4\pi}}{\Dp\Dct}\sum_{\lpr=0}^\infty\sum_{\mpr=-\lpr}^{+\lpr}A_{\lpr,\mpr}(Q)\times \nonumber \\
& \qquad    \int_{\phi_i-\Dp/2}^{\phi_i+\Dp/2}d\phi 
    \int_{\cti-\Dct/2}^{\cti+\Dct/2}d(\cos\theta)
    Y^*_{\lpr,\mpr}(\ct,\phi) \nonumber\\
& = 
    \sqrt{4\pi}\sum_{\lpr=0}^\infty\sum_{\mpr=-\lpr}^{+\lpr}A_{\lpr,\mpr}(Q)\cdot F_{\lpr,\mpr}(\Dp,\Dct,\cti)\times \nonumber\\
& \qquad  Y^*_{\lpr,\mpr}(\cti,\phi_i) .
\end{align}

     Here,
\begin{align}
& F_{\lpr,\mpr}(\Dp,\Dct,\cti)= \frac{\sin(m\Dp/2)}{m\Dp/2}\times \\
& \quad 
 \frac{1}{\Dct P_{\lpr,\mpr}(\cti)}\int_{\cti-\Dct/2}^{\cti+\Dct/2}d(\cos\theta) P_{\lpr,\mpr}(\ct) \nonumber
\end{align}
is the term which includes the finite binning effects.

Assuming that \Alm's vanish for $l,m$ greater than the sampling Nyquist frequency,
by the sampling theorem~\cite{Shannon:1948xx,Jerri:1977xx},
the \Alm's are completely determined by $C^\Delta$.
In fact, if $F_{l,m}$ were independent of $\cos\theta_i$, then we would have
\begin{align}
\AlmQ = &\frac{\Delta_\phi \Delta_{\cos\theta}}{F_{l,m}\left(\Dct,\Dp\right)\sqrt{4\pi}} \times \nonumber \\
&\sum_{{\rm bins} \thinspace i} C^\Delta\left(Q,\cti,\phi_i \right) Y_{l,m}\left(\cti,\phi_i\right)  , \nonumber
\end{align}
where the summation is over all bins of $\ct$ and $\phi$, for a given $Q$.

However, $F_{l,m}$ does depend on $\cti$, so the above equation does not strictly hold.
Nevertheless, we find, numerically, that an excellent approximation is 
\begin{equation}
\label{eq:ApproxBinningCorrection}
\AlmQ \approx \frac{\Delta_\phi \Delta_{\cos\theta}}{\sqrt{4\pi}} 
\sum_{{\rm bins} \thinspace i} \frac{C^\Delta\left(Q,\cti,\phi_i \right) Y_{l,m}\left(\cti,\phi_i\right)}{F_{l,m}\left(\Dct,\Dp,\cti\right)}  . 
\end{equation}

For any given measurement, one may check the validity of this approximation by plugging the result of Equation~\ref{eq:ApproxBinningCorrection}
into the expression on the last line of Equation~\ref{eq:binning}.  To the extent that it returns the measured correlation function $C^\Delta$,
the \Alm's returned by Equation~\ref{eq:ApproxBinningCorrection} are correctly extracted.  If there are deviations, the correct \Alm's can be
found by iterative techniques.

Other methods to remove binning effects have also been proposed~\cite{DaveBrownCoRALtoBePublished}.

\bibliographystyle{annrev}

\end{document}